\documentclass[%
reprint,
showkeys,
showpacs,preprintnumbers,
amsmath,amssymb,
aps,
pra,
floatfix,
twocolumn]{revtex4-2}
\usepackage{appendix}
\usepackage{amsmath,latexsym}
\usepackage{scalerel}
\usepackage{hyperref}
\usepackage{mathtools}
\usepackage{newtxtext,newtxmath}
\usepackage{setspace}
\RequirePackage{stackengine}
\RequirePackage{float}
\usepackage[caption=false]{subfig}
 
\begin{document}
	\title{Tsallis holographic dark energy reconsidered}
	\author{M. Dheepika}
	\email{mdheepika@cusat.ac.in}
	\author{Titus K. Mathew}
	\email{titus@cusat.ac.in}
	\affiliation{Department of Physics, Cochin University of Science and Technology, Kochi, Kerala 682022, India}
	\begin{abstract}{\vspace{1.0cm}\textbf{Abstract.}}
			We consider the interacting Tsallis Holographic Dark Energy (THDE), with the Granda-Oliveros (GO) scale as the infrared (IR) cutoff, as dynamical vacuum. We analytically solved for the Hubble parameter, in a spatially flat FLRW universe with dark energy and matter as components, and the solution traces the evolutionary path from the prior decelerated to the late accelerated epoch. Without interaction, the model predicts a $\Lambda$CDM like behavior with an effective cosmological constant. We used Pantheon Supernovae type Ia, observational Hubble data (OHD), cosmic microwave background (CMB), and baryon acoustic oscillation (BAO) data to constrain the free parameters of the model. The estimated values of the cosmological parameters were consistent with observational results. We analyzed the behavior of the model using the statefinder and $\omega^\prime_{e}-\omega_{e}$ plane where $\omega_{e}$ and $\omega^\prime_{e}$ corresponds to the effective equation of state and its evolution, respectively. The model shows a quintessence behavior in general, and the model trajectory ends in a point that corresponds to the de Sitter phase. We performed a dynamical analysis of the model, concluding that the prior decelerated and late accelerated phases are unstable and stable equilibria, respectively. We also investigated the thermodynamical nature of the model and found that the generalized second law remains valid in the dynamical vacuum treatment of the model.
		
	\end{abstract}

	\maketitle
	
	\newpage
\section{Introduction}
\label{intro}
After the discovery of the accelerated expansion of the universe \cite{SupernovaSearchTeam:1998fmf,SupernovaCosmologyProject:1998vns}, the desire to understand the universe spired, resulting in intensive works in recent literature. The nature and origin of the accelerated expansion of the universe are a mystery even now. Postulating dark energy models and modified theories of gravity are the two approaches that try to unwind the mystery of the universe's accelerated expansion. One of the simplest and best dark energy candidates is the cosmological constant, $\Lambda$ as given in the most successful model, $\Lambda$CDM. However, this model suffers mainly from two problems; one is the cosmological constant problem where there is a discrepancy in the observational and the theoretical value ($10^{121}$ times larger) of the energy density of the cosmological constant. The other is the coincidence problem that is the unknown reason for the coincidence of matter density and the cosmological constant during the current epoch, despite their different evolutionary nature. Dynamical dark energy models were suggested to mitigate these problems. Some representatives of this category are quintessence \cite{Caldwell:1997ii, zlatev1999quintessence}, k-essence \cite{Armendariz-Picon:2000nqq},  and phantom \cite{Caldwell:1999ew} models.	

One of the dynamical dark energy models is the holographic dark energy (HDE) model, which pivots on the black hole thermodynamics, proposed mainly to allay the coincidence problem. The underlying idea of these models is the holographic principle \cite{tHooft:1993dmi,Susskind:1994vu} which states that the entropy of a gravitating system is related to its surface area, not its volume, which implies that the maximum entropy of any region of space should be less than or equal to the entropy of a black hole of similar size. The limit set by this principle implies a relation between the short distance (ultraviolet-UV) cutoff and the long distance (infrared-IR) cutoff \cite{PhysRevLett.82.4971}. This idea reconciles the breakdown of the quantum field theory to describe a black hole \cite{PhysRevLett.82.4971, 2005PhLB..610...18M}. The initial proposal of the holographic principle considers the entropy of the cosmological horizon as the Bekenstein-Hawking entropy and the dark energy density scales as the square of the Hubble parameter. However, it could not explain the current accelerated expansion of the universe \cite{pavon2007holographic,Granda:2008dk}. HDE model with particle horizon \cite{fischler1998holography,li2004model} as IR cutoff also could not explain the present acceleration. Taking future event horizon \cite{li2004model} as IR cutoff successfully explained the present accelerated expansion, but it suffered from causality problem. Since any known symmetry does not dodge the interaction between the dark sectors \cite{wetterich1994cosmon}, models considering such interactions were proposed, and they yield better consistency with the cosmic observations than non interacting models  \cite{bertolami2007dark,olivares2005observational,pavon2005holographic, zimdahl2007interacting}. 

Holographic Ricci dark energy model with Ricci scalar curvature as IR cutoff introduced in \cite{2009PhRvD..79d3511G} is a phenomenologically viable model which avoids causality and coincidence problems. Inspired by this model and on a pure dimensional basis, Granda and Oliveros proposed a new IR cutoff \cite{Granda:2008dk}, a combined function of Hubble parameter and its time derivative. The resulting model also successfully avoids the coincidence problem, causality problem and explains the current accelerated expansion. Further explorations in HDE models with different IR cutoffs can be found in \cite{Nojiri:2005pu,PhysRevD.84.123507,2012PhRvD..85l7301C,2013EPJC...73.2352C,2017EPJC...77..528N}. In order to incorporate the quantum corrections 
\cite{das2002general,das2008black,das2008power,Das:2010su,radicella2010generalized,das2012entanglement,Tsallis:1987eu,Lyra:1997ggy,tsallis1998role, wilk2000interpretation}, conventional HDE required modifications. 

Tsallis and Cirto \cite{Tsallis:2012js} introduced a generalized non additive entropy, popularly known as Tsallis entropy, to solve the thermodynamic inconsistencies in non standard systems like a black hole. The pioneer works on the analysis of dark energy models with Tsallis non extensive statistical formulation can be found in \cite{2014arXiv1403.5706N} and further possibilities in cosmology are probed in \cite{2016JCAP...08..051N}. This kind of entropy agrees well with the Friedmann equations and Padmanabhan's proposal of the emergence of space time \cite{moradpour2016implications}. Like in the conventional HDE model, it is possible to construct dark energy models using Tsallis entropy, and as a result, Tsallis holographic dark energy (THDE) with Hubble horizon as IR cutoff was introduced in \cite{tavayef2018tsallis}. Taking inspiration from the aforementioned study,  dynamics of FRW universe having dark matter and THDE with the apparent horizon, the particle horizon, the Ricci scalar curvature scale, and the Granda-Oliveros (GO) scale as IR cutoffs was studied considering non interacting and interacting scenarios \cite{zadeh2018note,Aly2019TsallisHD,Srivastava:2020hng,Sharma:2021dqj}. It is found that the THDE model with particle horizon as IR cutoff explains the current accelerated expansion of the universe, unlike the corresponding conventional HDE model. The results from \cite{zadeh2018note} show that the THDE model is not always stable for the GO scale and the Ricci scalar cutoffs in both interacting and non interacting cases. Whereas in \cite{Aly2019TsallisHD} THDE model with the GO scale as IR cutoff shows stability in $(n+1)$ dimensional FRW universe. Thermodynamical stability studies of THDE with the apparent horizon as IR cutoff in \cite{AbdollahiZadeh2019ThermalSO} shows that the model does not satisfy the stability conditions in both interacting and non interacting cases. The investigations on the evolution of the THDE with Hubble horizon as IR cutoff, by considering time varying deceleration parameter in FRW universe is discussed in \cite{DIXIT2019101281}, in Brans-Dicke cosmology is discussed in \cite{aditya2019observational,Yadav:2020wsd}. Geometrical diagnosis of  THDE model of the universe with the apparent horizon as IR cutoff, considering the interaction between dark sectors of the universe, was made in \cite{Sharma:2019bgp}. Cosmological model in higher dimensional Kaluza-Klien theory, having THDE with Hubble horizon as IR cutoff, and with Generalized Chaplygin Gas (GCG)  as cosmic components are studied in \cite{Saha:2020vxn}. THDE with Hubble horizon as IR cutoff in Rastall framework and on Randall-Sundrum brane has been considered in \cite{ghaffari2020holographic,astashenok2020some}. Dynamical system studies on interacting and non interacting THDE in a fractal universe with Hubble radius and apparent horizon as IR cutoff can be found in \cite{Ghaffari:2019qcv,al2020study, al2020generalized,jawad2021cosmic}. The equivalence between Tsallis entropic dark energy and generalized HDE with cutoffs in terms of particle horizon, future horizon, and its derivatives are established in \cite{Nojiri:2021iko}. Cosmological analysis of the THDE with Hubble horizon as IR cutoff in the axially symmetric Bianchi-I universe within the framework of general relativity has been explained in \cite{dubey2019tsallis,ChandraDubey:2020tng}. Sign changeable mutual interactions between dark sectors are also considered to study the effects of anisotropy in the Bianchi universe \cite{AbdollahiZadeh:2019cqi}. Similar analysis of THDE with Hubble horizon and GO scale as IR cutoff in Bianchi-III universe has been discussed in \cite{sharma2020swampland,Korunur:2019rhg}.  Investigations on dynamics of THDE with Hubble horizon as IR cutoff, by assuming power law-exponential form for the scale factor have been studied in \cite{Bhattacharjee_2020}. Geometrical evolutionary studies in THDE models with Hubble horizon, future event horizon, and GO scale as IR cutoffs corresponding to different interactions have been explored in \cite{huang2019stability,Varshney2019fzj,sharma2019diagnosing,jawad2019non,zhang2020diagnosing,srivastava2020statefinder,IQBAL2019100349}. Investigations on THDE with GO scale as IR cutoff \cite{mohammadi2021tsallis}, presuming that this energy density is responsible for inflation, show the potentiality of the model in explaining the early universe. Comparison of THDE model with other HDE models using statefinder analysis was reviewed in \cite{dubey2021comparing}. The evolution of cosmological perturbations in THDE models with Hubble horizon and future event horizon as IR cutoff and Bayesian model comparison with $\Lambda$CDM as reference model has been scrutinized in \cite{d2019holographic,younas2019cosmological,da2021cosmological}. In \cite{aly2019study,sym11010092,Jawad:2019ouc,shaikh2021diagnosing,ens2020f,jawad2020generalized,jawad2020generalizedT,santhi2020bianchi,aly2020cosmological,rani2019cosmological,Shekh:2021bgh,Varshney:2020eun,sharma2020reconstruction,Varshney:2021xvg,VIJAYASANTHI2021101648,maity2020study,Liu:2021heo,Zubair:2021yrq}, the THDE model with different IR cutoffs within  various modified gravity theories and scalar field theories has also been explored. The growth rate of clustering for different IR cutoffs for the THDE model in the FRW universe can be found in \cite{bhattacharjee2021growth}. Cosmological implications through non linear interactions between THDE with Hubble horizon as IR cutoff and cold dark matter in the framework of loop quantum cosmology has been discussed in \cite{sym10110635}. Investigations on the THDE model with Hubble horizon as IR cutoff in the higher derivative theory of gravity in  \cite{pradhan2021tsallis} show that it is not compatible with late time acceleration as it could not acquire the required value of the equation of state parameter.  Due to the quantified non extensivity in Tsallis entropy, studies on modification of Friedmann equation and gravity theory, including emergence proposal of gravity \cite{lymperis2018modified,sheykhi2018modified,nojiri2019modified,nojiri2020correspondence,abbasi2020tsallisian,asghari2021observational}, also flourish in this field. All these works were carried out considering the equation of state parameter of dark energy as varying with the expansion of the universe.

It is to be noted that, in formulating the HDE density, one has to compare the UV cutoff, corresponding to the vacuum energy, with the IR cutoff, representing the large length scale of the universe. There is a broad consensus that the cosmological constant in the standard $\Lambda$CDM model can be the vacuum energy, having an equation of state parameter, $-1$. Hence, the UV cutoff involved in the HDE models indicates that the corresponding dark energy density is of dynamical vacuum. As a result, reconsidering the THDE as the dynamical vacuum in analyzing the evolutional history of the universe is of practical significance. We are considering the THDE as the dynamical vacuum in the present work. We study the evolution of the universe by considering the interaction, between dark matter and dark energy in conformity with the total conservation of energy, in the THDE model with the GO scale as IR cutoff. Our analysis shows that the model predicts a transition to the late accelerating universe. This work also involves the geometrical and dynamical analysis and thermodynamical study of the model to check the feasibility of explaining the accelerating universe.  

The structure of this paper is as follows. In the next section, we present an interacting THDE as a dynamical vacuum. We analytically solve for the Hubble parameter and investigate its evolutionary behavior. In Sect.~\ref{sec:2}, we constrain the parameters with observational data and discuss its cosmological implications. Along with that, we also analyze the evolutionary trajectory of the model in geometrical plane $r-s$ and phase plane $\omega^\prime_{e}-\omega_{e}$ plane. In Sect.~\ref{sec:3}, we perform the dynamical analysis on the interacting THDE model. In Sect.~\ref{sec:4}, we study the thermodynamical properties of the model. In the last section, we summarize the conclusions of the work.

\section{Interacting THDE model as dynamical vacuum}
\label{sec:1}
A generalization of the Boltzmann-Gibbs (BG) theory, now known as the non extensive statistical mechanics, was proposed \cite{Tsallis:1987eu} to address the complexities in non standard systems. For large scale systems, the thermodynamical entropy must be modified to non additive entropy \cite{Tsallis:2012js}. According to Tsallis and Cirto \cite{Tsallis:2012js}, the quantum correction modified the entropy area relation as,
\begin{equation}\label{eqn:S1:1}
	S = \gamma A^{\delta},
\end{equation}
where $A$ is the horizon area of the black hole, $\gamma$ is a positive \cite{lymperis2018modified} constant, and $\delta$ is the positive non additive parameter
\cite{Tsallis:2012js}. This will reduces to the Bekenstein entropy for $\gamma = \frac{1}{4L_{p}^{2}}$ and $\delta=1,$ with $L_{p}^{2}$ as the Planck length. Following the holographic principle, Cohen et al. \cite{PhysRevLett.82.4971} have found  a relation between the entropy, IR cutoff (L), and the UV cutoff ($\Lambda$) as,
\begin{equation}\label{eqn:S1:2}
	L^3 \Lambda^3 \leq S^{3/4}.
\end{equation}
Following the Tsallis entropy in \eqref{eqn:S1:1} and substituting for area, $A=4\pi L^2,$ leads to the relation $\Lambda^4 \leq \gamma (4\pi)^{\delta} L^{2\delta -4},$ and it gives a measure of the vacuum energy density. Taking consideration of the equality in this relation, a modified energy density, known as the Tsallis HDE (THDE) density, can be defined as,
\begin{equation}\label{eqn:S1:3}
	\rho_{de} = C L^{2\delta -4},
\end{equation}
where the constant, $C=\gamma (4\pi)^{\delta}$ with dimension [$L^{2-2\delta}$] (in units of $8\pi G= \hbar =c =1$). The simplest choice for scale is $L=H^{-1},$  the Hubble horizon.
In \cite{tavayef2018tsallis}, authors have established that the corresponding model of the universe, with non interacting dark sectors, can show a transition into the late time accelerated epoch. This contrasts with the conventional HDE model, which failed to predict a transition into the late accelerated epoch if cosmic components are non interacting.

In the present study we adopt the GO scale as IR cutoff, which was originally proposed in reference \cite{Granda:2008dk} to study the conventional HDE model and is given by,
\begin{equation}\label{eqn:S1:4}
	L^{-2}=(\alpha H^{2}+\beta\dot{H}),
\end{equation}
where $\alpha$ and $\beta$ are unknown dimensionless constants and $\dot{H}$, is the derivative of Hubble parameter with respect to the cosmic time.
Using \eqref{eqn:S1:4} in \eqref{eqn:S1:3}, THDE density can be written as
\begin{equation}\label{eqn:S1:5}
	\rho_{de}=3(\alpha^\prime H^{2}+\beta^\prime \dot{H})^{2-\delta},
\end{equation}
where $\alpha^\prime=\frac{\alpha}{3}C^{\frac{1}{2-\delta}}$ and $\beta^\prime=\frac{\beta}{3}C^{\frac{1}{2-\delta}}$ have dimension $[L^{\frac{2-2\delta}{2-\delta}}]$.

The Friedmann equation for the flat FRW universe is given by
\begin{equation}\label{eqn:S1:6}
	3H^{2}=\rho_{m}+\rho_{de}.
\end{equation}
where $\rho_{m}$ and $\rho_{de}$ is the dark matter and dark energy density respectively.
The conservation equations including the interaction between THDE and dark matter are given by
\begin{equation}\label{eqn:S1:7}
	\dot{\rho}_{de}+3H(\rho_{de}+P_{de})=-Q,~~~ \dot{\rho}_{m}+3H(\rho_{m}+P_{m})=Q,
\end{equation}
where $\dot{\rho}_{de}$ and $\dot{\rho}_{m}$ are the derivatives of dark energy and dark matter densities with respect to the cosmic time, $P_{de}$ and $P_{m}$ are the pressure of dark energy and matter respectively and $Q$ represents the interaction, which determines the rate of exchange of energy between the dark sectors. From equation \eqref{eqn:S1:7} it is clear that $Q$ has to be a function of energy density and inverse of time. We are adopting a simple function, $Q=3bH\rho_{m}$ where $b$ is the coupling constant. None of the previous works has considered this form of interaction in combination with the GO scale in studying the evolution of the FLRW universe. Since the THDE is considered as a dynamical vacuum, its equation of state is $P_{de}=-\rho_{de}$ and the matter is considered as pressureless. Considering the above assumptions, the equations in \eqref{eqn:S1:7} reduces to
\begin{equation}\label{eqn:S1:9}
	\dot{\rho}_{de}=-3bH\rho_{m}, ~~~ \dot{\rho_{m}}=-3(1-b)H\rho_{m}.
\end{equation}
The above equations \eqref{eqn:S1:9} can be rewritten in terms of density parameter $\Omega=\frac{\rho}{3H_{0}^{2}}$ where $H_{0}$ denotes the present value of Hubble parameter and $x=\ln a$ as
\begin{equation}\label{eqn:S1:11}
	\frac{d\Omega_{de}}{dx}=-3b\Omega_{m},
\end{equation}
\begin{equation}\label{eqn:S1:12}
	\frac{d\Omega_{m}}{dx}=-3(1-b)\Omega_{m}.	
\end{equation}
The solution of equation \eqref{eqn:S1:12} is  $\Omega_{m}=\Omega_{m0}e^{-3(1-b)x}$ where $\Omega_{m0}=\frac{\rho_{m0}}{3H_{0}^{2}}$ is the present matter density parameter. In similar manner, $\Omega_{de0}=\frac{\rho_{de0}}{3H_{0}^{2}}$ is the present dark energy density parameter. The equation \eqref{6} reduces to, $1=\Omega_{m0}+\Omega_{de0}$, for $H=H_{0}$. Considering this result and using equations \eqref{eqn:S1:12}, \eqref{eqn:S1:11} and \eqref{eqn:S1:6}, a second order differential equation can be formulated as below
\begin{equation}\label{eqn:S1:13}
	\frac{d^{2}h^2}{dx^2}+3\dfrac{dh^2}{dx}+9b\Omega_{m0}e^{-3(1-b)x}=0.
\end{equation}
where $h^2=H^{2}/H_{0}^{2}$. The solution of the above differential equation in terms of scale factor is \cite{George:2018myt},
\begin{equation}\label{eqn:S1:14}
	h^2=
	\frac{\Omega_{m0}}{1-b}a^{-3(1-b)}+\frac{1}{3}c^\prime a^{-3}+c^{\prime\prime},
\end{equation}
which tells us how the Hubble parameter evolves with respect to the scale factor of the universe. In the previous works of THDE model with GO scale as IR cutoff \cite{zadeh2018note,zhang2020diagnosing,bhattacharjee2021growth}, such an exact solution has not been obtained.
Initial conditions to evaluate the constants $c^{\prime}$ and $c^{\prime\prime}$ are  
\begin{equation}\label{eqn:S1:15}
	h^2\vert_{a=1}=1,~~~~ \frac{dh^2}{dx}\bigg\vert_{a=1}=
	\dfrac{2}{\beta^{\prime}}\left[\left(\frac{\Omega_{de0}}{H_{0}^{2(1-\delta)}}\right)^{\dfrac{1}{2-\delta}}-\alpha^{\prime}\right].
\end{equation}
The constants are then obtained as
\begin{equation}\label{eqn:S1:16}
	c^\prime=\frac{2}{\beta^{\prime}}\left[\alpha^{\prime} - \left(\frac{\Omega_{de0}}{H_{0}^{2(1-\delta)}}\right)^\frac{1}{2-\delta}\right] - 3\Omega_{m0}, 
\end{equation}
\begin{equation}\label{eqn:S1:17}
	c^{\prime\prime}=1-\frac{b\Omega_{m0}}{1-b}-\frac{2}{3\beta^{\prime}}\left[\alpha^{\prime} - \left(\frac{\Omega_{de0}}{H_{0}^{2(1-\delta)}}\right)^\frac{1}{2-\delta}\right].
\end{equation}
In the asymptotic limit $a \to 0$, the constant $c^{\prime\prime}$ can be neglected due to domination of the first two terms in the Hubble parameter equation \eqref{eqn:S1:14} for $b\ll1$, consequently the resulting solution represents the decelerated expansion. In the future limit, $a \to \infty$, the constant term in the Hubble parameter will dominate over the rest of the terms,  indicating an end de Sitter phase. Hence the model predicts a transition into a late accelerating epoch in the evolution of the universe.   
For $b=0$ equation \eqref{eqn:S1:14} will reduce to
\begin{equation}\label{hbzero}
	h^2 = \tilde{\Omega}_{m0} a^{-3} + \tilde{\Omega}_{de0} ,
\end{equation}
where $\tilde{\Omega}_{m0}$ and $ \tilde{\Omega}_{de0}$ are the mass density parameters \cite{Starobinsky:1998fr} which have the form
\begin{equation}
	\tilde{\Omega}_{m0}= \frac{2}{3\beta^{\prime}} \left[\alpha^{\prime}-\left(\frac{1-\Omega_{m0}}{H_0^{2(1-\delta)}}\right)^{\frac{1}{2-\delta}}\right], \quad \tilde{\Omega}_{de0} = 1-\tilde{\Omega}_{m0}. 
\end{equation}
This shows that, in the absence of interaction, the present model is similar to that of the standard $\Lambda$CDM, with an effective cosmological constant corresponding to the mass density parameter $ \tilde{\Omega}_{de0}$.  Even though the dark energy density in equation (\ref{eqn:S1:5}) is varying with $H,$ the behavior of it as effective vacuum energy with equation of state, $\omega_{de}=-1$ is the primary reason for this $\Lambda$CDM like behavior, in the absence of the interaction. Further, the coupled conservation equations in \eqref{eqn:S1:7} become independent conservation, in which matter and dark energy are separately conserved. Under such a condition, the dark energy density, equivalent to the vacuum energy density, will effectively become a constant. In contrast, the previous works on the non interacting THDE models \cite{tavayef2018tsallis,saridakis2018holographic,zadeh2018note,AbdollahiZadeh2019ThermalSO,Aly2019TsallisHD,sharma2019diagnosing,d2019holographic,DIXIT2019101281,Dubey:2019kzh,ebrahimi2020dynamical,srivastava2020statefinder,Sharma:2021dqj,dubey2021comparing} shows quintessence like, phantom like, or phantom divide crossing behavior according to the value of model parameters.
\section{Observational constraints and its cosmological implications}
\label{sec:2}
\subsection{The model parameter estimation}
 \begin{table*}
	\caption{The best estimated values and 68.3\% confidence limit for interacting THDE model parameters}
	\label{tab:1}
	\begin{tabular*}{\textwidth}{@{\extracolsep{\fill}}lrrrrrrl@{}}
		\hline\noalign{\smallskip}
		Data & \multicolumn{1}{c}{$\delta$} & \multicolumn{1}{c}{$b$} & \multicolumn{1}{c}{$\alpha^\prime$} & \multicolumn{1}{c}{$\beta^\prime$} & \multicolumn{1}{c}{$\Omega_{m0}$} & \multicolumn{1}{c}{$ H_{0}$}  & M \\
		\noalign{\smallskip}\hline\noalign{\smallskip}
		SNIa & $1.021^{+0.017}_{-0.020}$ & $-0.027^{+0.062}_{-0.054}$ & $0.963^{+0.139}_{-0.121}$ & $0.309^{+0.201}_{-0.185}$ & $0.300^{+0.035}_{-0.034}$ & $69.977^{+3.514}_{-3.436}$ & $-19.359^{+0.107}_{-0.109}$ \\	
		\noalign{\smallskip}\hline\noalign{\smallskip}
		OHD & $1.018^{+0.026}_{-0.031}$ & $0.039^{+0.133}_{-0.068}$ & $0.955^{+0.165}_{-0.166}$ & $0.305^{+0.212}_{-0.250}$ & $0.295^{+0.038}_{-0.032}$ & $69.506^{+2.705}_{-2.773}$ & $ \quad \quad \; - $ \\	
		\noalign{\smallskip}\hline\noalign{\smallskip}
		SNIa$+$OHD & $1.020^{+0.016}_{-0.021}$ & $0.023^{+0.033}_{-0.026}$ & $0.960^{+0.138}_{-0.125}$ & $0.337^{+0.184}_{-0.207}$ & $0.296^{+0.038}_{-0.033}$ & $69.241^{+1.230}_{-1.229}$ & $-19.381^{+0.037}_{-0.037}$ \\	
		\noalign{\smallskip}\hline\noalign{\smallskip}
		SNIa$+$OHD$+$CMB & $1.021^{+0.012}_{-0.014}$ & $0.014^{+0.016}_{-0.022}$ & $0.970^{+0.129}_{-0.094}$ & $0.256^{+0.227}_{-0.158}$ & $0.271^{+0.022}_{-0.015}$ & $68.867^{+1.199}_{-1.159}$ & $-19.393^{+0.035}_{-0.034}$ \\	
		\noalign{\smallskip}\hline\noalign{\smallskip}
		SNIa$+$OHD$+$BAO & $1.020^{+0.016}_{-0.022}$ & $0.028^{+0.033}_{-0.029}$ & $0.974^{+0.142}_{-0.132}$ & $0.352^{+0.172}_{-0.205}$ & $0.294^{+0.020}_{-0.019}$ & $69.267^{+1.209}_{-1.204}$ & $-19.379^{+0.036}_{-0.037}$ \\	
		\noalign{\smallskip}\hline\noalign{\smallskip}
		SNIa$+$OHD$+$CMB$+$BAO &   $1.021^{+0.012}_{-0.015}$& $0.005^{+0.016 }_{ -0.017 }$ & $0.981^{+0.128}_{-0.099}$ & $0.308^{+0.201}_{-0.181}$  & $0.281^{+0.017}_{-0.015}$ & $68.672^{+1.196}_{-1.141}$ & $-19.399^{+0.034}_{-0.034}$ \\	
		\noalign{\smallskip}\hline
	\end{tabular*}
\end{table*}
\begin{figure*}
	\subfloat[\label{gtc_shc}]{%
		\includegraphics[width=0.49\textwidth]{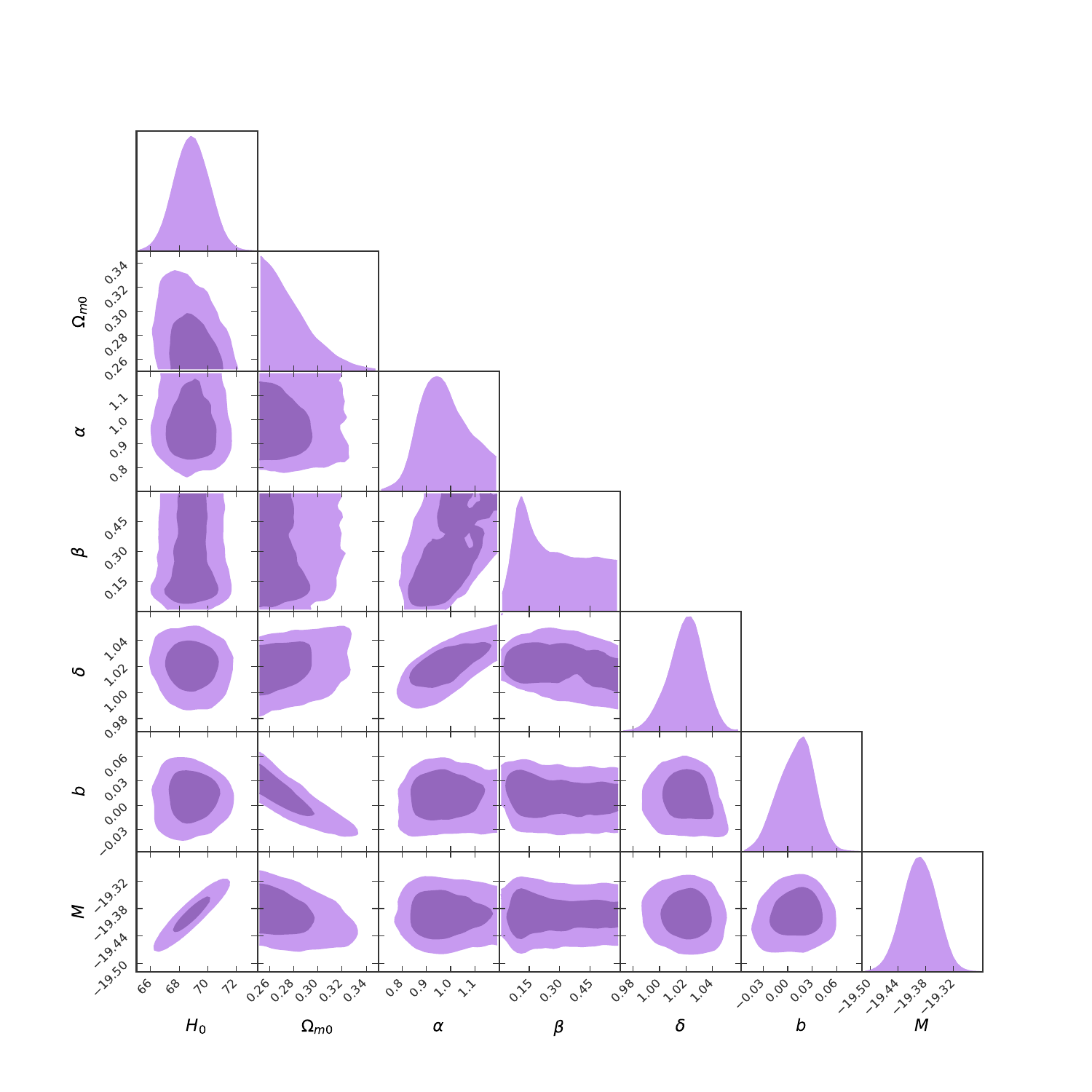}%
	}\hfill
	\subfloat[\label{gtc_cd}]{%
		\includegraphics[width=0.49\textwidth]{fig1.pdf}%
	}
	\caption{The two dimensional posterior contours with 68\% and 95\% confidence limits and one dimensional posterior distribution (using pygtc open python package \cite{Bocquet2016}) from the (a) SNIa$+$OHD$+$CMB and (b) SNIa$+$OHD$+$CMB$+$BAO datasets for the free parameters of THDE model.}
	\label{}
\end{figure*}
In this section, we estimate the constant model parameters,
$\delta$, $\alpha^\prime$, $\beta^\prime$, $b$, $H_0$ and $\Omega_{m0}$ by contrasting the model with cosmological observational data. The dataset consists of the type Ia supernovae data \cite{Scolnic:2017caz}, observational Hubble data (OHD) \cite{Yu:2017iju,Amirhashchi:2018nxl}, cosmic microwave background (CMB) data \cite{Chen:20db18v}, and baryon acoustic oscillation (BAO) data \cite{Blake:2011en}. We have applied the  Markov chain Monte Carlo (MCMC) method by employing the emcee python package \cite{foreman2013emcee} using the lmfit python library \cite{newville2021lmfit} to constrain the model parameters. To perform the analysis, we assume uniform priors for parameters specific to the model as follows: $\delta \in (0.1,5.0)$, $b \in (-0.1,0.5)$, $\alpha^\prime \in (0.7,1.2)$, and $\beta^\prime \in (0.001,0.6)$, considering the recent literature \cite{tavayef2018tsallis,saridakis2018holographic,zadeh2018note,AbdollahiZadeh2019ThermalSO,Aly2019TsallisHD,sharma2019diagnosing,Varshney2019fzj,DIXIT2019101281,huang2019stability,sadri2019observational,Dubey:2019kzh,zhang2020diagnosing,Sharma:2019bgp,ebrahimi2020dynamical,srivastava2020statefinder,Sharma:2021dqj,al2020generalized,dubey2021comparing,d2019holographic,da2021cosmological,bhattacharjee2021growth}.We have considered uniform priors from the recent literature suitably for other general parameters like $H_0$, $\Omega_{m0}$, and $M$.

We use the latest supernovae data, the Pantheon Sample, which comprises 1048 SNe Ia data points in the redshift range of $0.01<z<2.3$.
The luminosity distance of SN Ia can be obtained using the relation,
	\begin{align}\label{eqn:S2:1}
		d_{L}&(\delta,\alpha^\prime,\beta^\prime,b,H_{0},\Omega_{m0},z_{i})\nonumber\\
		&=c(1+z_{i})\int_{0}^{z_{i}}\frac{dz^\prime}{H(\delta,\alpha^\prime,\beta^\prime,b,H_{0},\Omega_{m0},z^\prime)},
	\end{align}
where $z_{i}$ is the redshift of the SN Ia, $c$ is the speed of light and $H(\delta,\alpha^\prime,\beta^\prime,b,H_{0},\Omega_{m0},z^\prime)$ is the Hubble parameter in terms of model parameters and redshift.

The theoretical distance modulus of SN Ia is given by 
	\begin{align}\label{eqn:S2:2}
		\mu_{th}&(\delta,\alpha^\prime,\beta^\prime,b,H_{0},\Omega_{m0},z_{i})\nonumber\\
		&=5\log_{10}\left[\frac{d_{L}(\delta,\alpha^\prime,\beta^\prime,b,H_{0},\Omega_{m0},z_{i})}{Mpc}\right]+25.
	\end{align}

The $\chi^{2}$ function of SN Ia data can be expressed as 
	\begin{align}\label{eqn:S2:3}
		\chi^{2}&(\delta,\alpha^\prime,\beta^\prime,b,H_{0},\Omega_{m0},M)_{SNIa}\nonumber\\
		&=\sum_{i=1}^{n} \frac{[\mu_{th}(\delta,\alpha^\prime,\beta^\prime,b,H_{0},\Omega_{m0},z_{i})-\mu_{i}]^{2}}{\sigma_{i}^{2}} ,
	\end{align}
where $\mu_{i}=m-M$ is the observational distance modulus of SN Ia, $m$ and $M$ are the apparent and the absolute magnitude of the SN Ia, $n=1048$, the total number of data points and $\sigma_{i}^{2}$ is the variance of $i^{th}$ measurement. 
We treat $M$ as a nuisance parameter throughout this analysis. The estimated values of the model parameters using the
SNIa data are given in the first row of  Table \ref{tab:1}. The minimum $\chi^{2}$ is obtained as $1035.485$.  The $\chi^{2}_{dof}=\frac{\chi^{2}_{min}}{n-n_{p}}$ is the minimum $\chi^{2}$ function per degrees of freedom where $n$ is the number of data points, and $n_{p}$ is the number of model parameters. The obtained value of $\chi^{2}$ goodness of fit, $\chi^{2}_{d.o.f.}=0.995$, i.e., around unity. The estimated value of the coupling constant is obtained to be negative, contributing a negative interaction term. It conveys the possibility of energy transfer from dark matter to THDE. 

Another independent observable we use is the Hubble dataset consisting of  $36$ $H(z)$ data points in the redshift range of $0.07\leq z \leq2.36$, out of which 31 data points are determined using the cosmic chronometric technique, 3 data points from the radial BAO signal in the galaxy distribution and 2 data points from the BAO signal in the Lyman $\alpha$ forest distribution alone or cross correlated with quasi stellar objects \cite{Yu:2017iju}. The following $\chi^{2}$ function is minimized for this Hubble parameter measurement
	\begin{align}\label{eqn:S2:4}
		\chi^{2}&(\delta,\alpha^\prime,\beta^\prime,b,H_{0},\Omega_{m0})_{OHD}\nonumber\\
		&=\sum_{i=1}^{n} \frac{[H(\delta,\alpha^\prime,\beta^\prime,b,H_{0},\Omega_{m0},z_{i})-H_{i}]^{2}}{\sigma_{i}^{2}} ,
	\end{align}
where $H_i$ is the observational Hubble parameter measurement, and $\sigma_{i}^{2}$ is the variance of $i^{th}$ measurement. The estimated values of the model parameters using the OHD data are given in the second row of  Table \ref{tab:1}. The minimum $\chi^{2}$ is obtained as $17.024$.  The obtained value of $\chi^{2}$ goodness of fit, $\chi^{2}_{d.o.f.}=0.567$.

Combining the type Ia supernovae data and the OHD, the total $\chi^{2}$ function takes the form
	\begin{align}
		\chi^{2}_{f}=&\chi^{2}(\delta,\alpha^\prime,\beta^\prime,b,H_{0},\Omega_{m0},M)_{SNIa}+\nonumber\\	&\chi^{2}(\delta,\alpha^\prime,\beta^\prime,b,H_{0},\Omega_{m0})_{OHD}.
	\end{align}

The estimated values of the model parameters using the SNIa data and OHD are given in the third row of  Table \ref{tab:1}. The minimum $\chi^{2}_{f}$ is obtained as $1071.352$. The obtained value of $\chi^{2}$ goodness of fit, $\chi^{2}_{d.o.f.}=0.995$, i.e., around unity.

The distance prior, shift parameter, $R$, which influences the CMB temperature power spectrum, is defined in terms of THDE model parameters as follows
	\begin{align}\label{eqn:S2:cmb1}
		R&(\delta,\alpha^\prime,\beta^\prime,b,H_{0},\Omega_{m0})\nonumber\\
		&=\sqrt{\Omega_{m0}}\int_{0}^{z_{\ast}}\frac{dz}{h(\delta,\alpha^\prime,\beta^\prime,b,H_{0},\Omega_{m0},z)},
	\end{align}
where $z_{\ast}$ is the redshift at the photon decoupling epoch. We adopt the distance prior measurement value $R_{obs}=1.7502\pm0.0046$ at the redshift $z_{\ast}=1089.92$ from the Planck 2018 observations \cite{Chen:20db18v}.  The corresponding $\chi^{2}$ \cite{Yin:2019rgm} from the CMB data is
	\begin{align}\label{eqn:S2:cmb2}
		\chi^{2}&(\delta,\alpha^\prime,\beta^\prime,b,H_{0},\Omega_{m0},M)_{CMB}\nonumber\\
		&= \frac{[R(\delta,\alpha^\prime,\beta^\prime,b,H_{0},\Omega_{m0})-R_{obs}]^{2}}{\sigma_{R}^{2}},
	\end{align}
where $\sigma_{R}$ is the variance of the $R_{obs}$ measurement. The estimated values of the model parameters using the SNIa$+$OHD$+$CMB are given in the fourth row of  Table \ref{tab:1}. The minimum $\chi^{2}$ is obtained as $1203.630$. The obtained value of $\chi^{2}$ goodness of fit, $\chi^{2}_{d.o.f.}=1.117$, i.e., around unity. 

The distance–redshift relation determined by a BAO measurement is given by the acoustic peak parameter, $A$, which is defined in terms of THDE model parameters as follows
	\begin{align}\label{eqn:S2:bao1}
		A&(\delta,\alpha^\prime,\beta^\prime,b,H_{0},\Omega_{m0})\nonumber\\
		&=\frac{\sqrt{\Omega_{m0}}}{h(z_A)^{1/3}}\left(\frac{1}{z_A}\int_{0}^{z_{A}}\frac{dz}{h(\delta,\alpha^\prime,\beta^\prime,b,H_{0},\Omega_{m0},z)}\right)^{\frac{2}{3}},
	\end{align}
where $z_A$ is the redshift of the acoustic peak parameter. We adopt the value of $A_{obs}=0.484\pm0.016$ at the redshift $z_{A}=0.35$ from the SDSS-BAO distance data \cite{blake2011wigglez}. The corresponding $\chi^{2}$ from the BAO data is
	\begin{align}\label{eqn:S2:bao2}
		\chi^{2}&(\delta,\alpha^\prime,\beta^\prime,b,H_{0},\Omega_{m0},M)_{BAO}\nonumber\\
		&= \frac{[A(\delta,\alpha^\prime,\beta^\prime,b,H_{0},\Omega_{m0})-A_{obs}]^{2}}{\sigma_{A}^{2}},
	\end{align}
where $\sigma_{A}$ is the variance of the $A_{obs}$ measurement. The estimated values of the model parameters using the SNIa$+$OHD$+$BAO are given in the fifth row of  Table \ref{tab:1}. The minimum $\chi^{2}$ is obtained as $1067.240$. The obtained value of $\chi^{2}$ goodness of fit, $\chi^{2}_{d.o.f.}=0.990$, i.e., around unity.

Combining the type Ia supernovae data, the OHD, the CMB, and the BAO data, the total $\chi^{2}$ function takes the form
	\begin{align}
		\chi^{2}_{s}=&\chi^{2}(\delta,\alpha^\prime,\beta^\prime,b,H_{0},\Omega_{m0},M)_{SNIa}+\nonumber\\	&\chi^{2}(\delta,\alpha^\prime,\beta^\prime,b,H_{0},\Omega_{m0})_{OHD}+\nonumber\\
		&\chi^{2}(\delta,\alpha^\prime,\beta^\prime,b,H_{0},\Omega_{m0},M)_{CMB}+\nonumber\\
		&\chi^{2}(\delta,\alpha^\prime,\beta^\prime,b,H_{0},\Omega_{m0},M)_{BAO}.
	\end{align}

The estimated values of the model parameters using the type Ia supernovae data, the OHD, the CMB, and the BAO data are given in the sixth row of Table \ref{tab:1}. The minimum $\chi^{2}_{s}$ is obtained as $1174.015$. The obtained value of $\chi^{2}$ goodness of fit, $\chi^{2}_{d.o.f.}=1.088$, i.e., around unity. The estimated value of the coupling constant is obtained to be positive and is smaller than the value obtained from all the other datasets except SNIa for the present THDE model and the previous THDE models \cite{sadri2019observational} with Hubble horizon and future event horizon as IR cutoff. The positive value of the coupling constant contributes to a positive interaction term, consequently the energy exchange is from THDE to dark matter and thereby satisfying the Le Chatelier-Braun principle \cite{pavon2009chatelier,al2020generalized}. The uncertainty on $b$ is relatively high compared to the value at which the maximum likelihood function peaks for all datasets. Evidence for such plausibility using SNIa and OHD can be observed in the work of C. P. Singh \cite{PhysRevD.102.123537} on Holographic dark energy. The best fit value of $\delta$ obtained using  all the datasets is similar to the previous results from THDE models with future event horizon as IR cutoff \cite{saridakis2018holographic,sadri2019observational}, which are stable against the background perturbations. Concurrently, it is smaller than the best fit value ($\delta>2$) obtained from the studies on the THDE model with Hubble horizon as IR cutoff \cite{sadri2019observational}, which is unstable against the background perturbation. Furthermore, it is greater than the best fit $\delta$ value (less than $1$) obtained from the study on the growth of matter fluctuations of the THDE model \cite{d2019holographic} with future event horizon as IR cutoff using the Gold-2017 dataset of 18 uncorrelated $\sigma_8$ measurements, SNIa data, and OHD.
\subsection{Evolution of cosmological parameters}
\begin{figure}[t!]
	\includegraphics[width=0.45\textwidth]{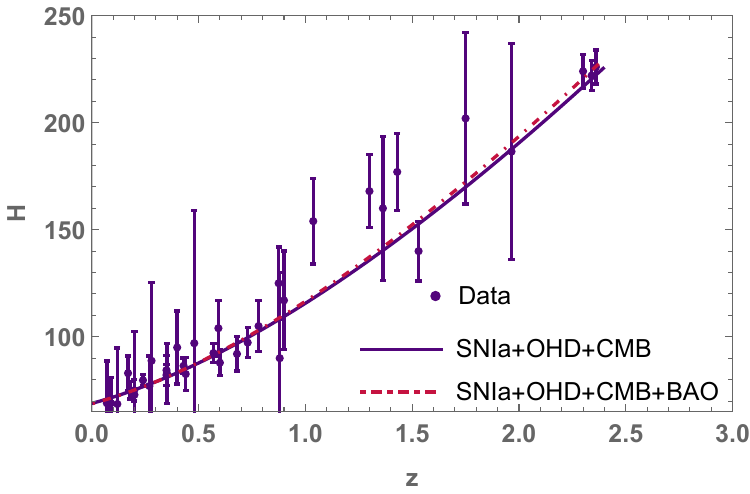} 
	\caption{The evolution of Hubble parameter with redshift for the best estimated values of the model parameters using the fourth and sixth datasets, compared with $H(z)$ data points
	}
	\label{HP_zplot} 
\end{figure}
\begin{figure}[t!]
	\includegraphics[width=0.45\textwidth]{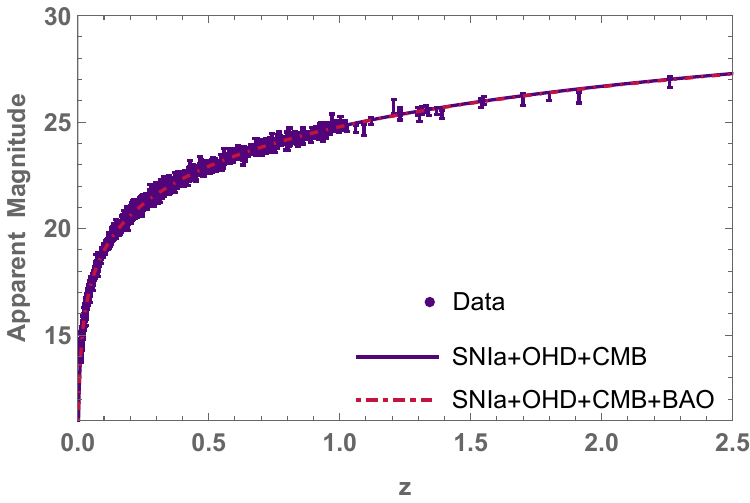} 
	\caption{The evolution of apparent magnitude with redshift for the best estimated values of the model parameters using the fourth and sixth datasets, compared with type Ia SNe data points}
	\label{DM_zplot} 
\end{figure}

Estimating luminosity and redshift from the cosmological observables contributes to determining the present Hubble parameter of the universe. Here, we studied the background evolution of the Hubble parameter using the interacting THDE solution. The predicted evolution of the Hubble parameter with redshift using this model has been compared with the 36 $H(z)$ data points and is shown in Fig. \ref{HP_zplot}. The Hubble parameter decreases as the cosmic time evolves and the present value of the Hubble parameter is obtained as $68.867^{+1.199}_{-1.159}$ kms$^{-1}$Mpc$^{-1}$ (fourth dataset) and $68.672^{+1.196}_{-1.141}$ kms$^{-1}$Mpc$^{-1}$ (sixth dataset) which is slightly lower than the observational value $H_{0}=70.5 \pm 1.3$  kms$^{-1}$Mpc$^{-1}$ from WMAP + BAO + SN data \cite{Hinshaw2008FiveYearWM} and slightly higher than observational value $H_{0}=67.37 \pm 0.54$ kms$^{-1}$Mpc$^{-1}$ from Planck data \cite{Planck:2018vyg} assuming $\Lambda$CDM model. The estimated present value of the Hubble parameter in our model is close to the best estimated values of Hubble parameter in interacting and non interacting THDE models with Hubble horizon and future event horizon as IR cutoff \cite{saridakis2018holographic,d2019holographic,sadri2019observational}. The theoretical evolution of apparent magnitude with respect to the redshift using this model is compared with the 1048 SNe Ia data points as shown in Fig. \ref{DM_zplot}. Both the error bar plots show proximity between theoretical and observational results.

\begin{figure}
	\def\big{
		\includegraphics[width=0.45\textwidth]{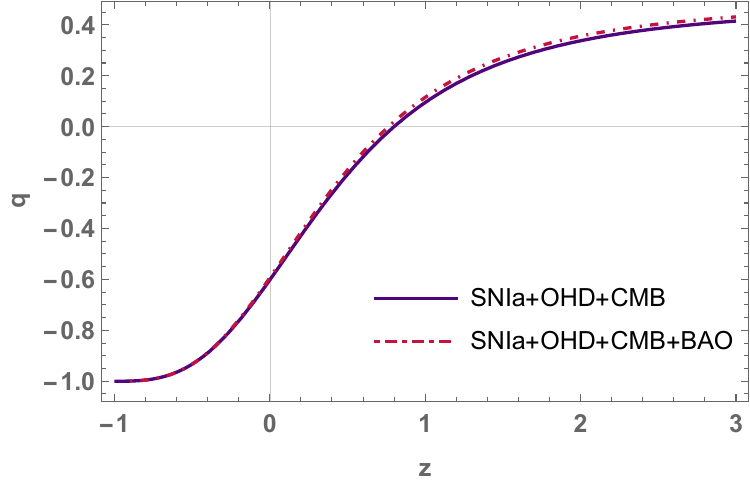}
	}
	\def\little{\includegraphics[height=1.4cm,width=1.8cm]{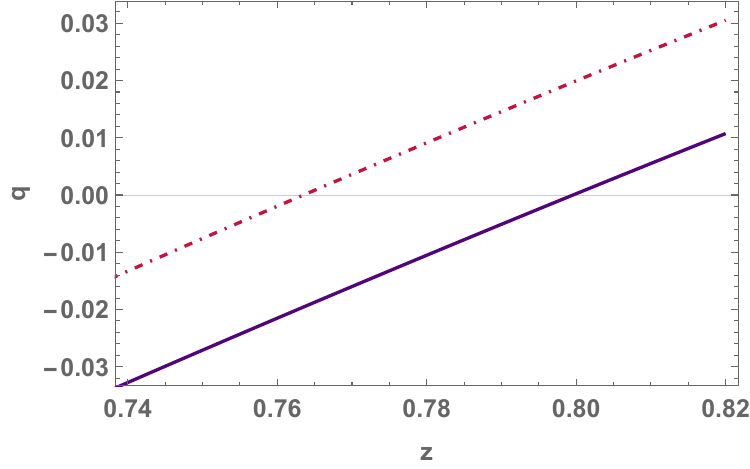}}
	
	\stackinset{l}{34pt}{t}{21.95pt}{\little}{\big}
	\caption{The evolution of deceleration parameter with redshift for the best estimated values of the model parameters using the fourth and sixth datasets. The inset figure clearly distinguishes the continuous and dot-dashed lines}	\label{q_zplot} 
\end{figure}

The deceleration parameter which measures the rate of cosmic expansion is obtained in the following form
\begin{equation}\label{eqn:S2.2:1}
	q=-1-\frac{\dot{H}}{H^{2}}= -1+\frac{3\Omega_{m0}(1+z)^{3(1-b)}+c^\prime (1+z)^{3}}{\dfrac{2\Omega_{m0}}{1-b}(1+z)^{3(1-b)}+\dfrac{2}{3}c^\prime (1+z)^{3}+2c^{\prime\prime}}.
\end{equation}
In the limit $z \to -1$, the deceleration parameter $q \to -1$ and in the limit $z \to \infty$, the deceleration parameter will tend to a positive value since second term in equation \eqref{eqn:S2.2:1} dominates in that case. The evolution of the deceleration parameter $q(z)$, as a function of z,  for the best estimated model parameters from the fourth and sixth datasets is plotted in Fig. \ref{q_zplot}. It is lucid from the figure that the model explains the current accelerated universe and also the transition from the prior decelerated phase to the present accelerated phase. The inset figure in Fig. \ref{q_zplot} shows the transition in small scale to reveal precisely the difference in the transition redshift for different datasets. The transition redshift $z_{t}$ $(i.e.,q(z_{t})=0)$ is found to be $0.799$ using the fourth dataset and $0.763$ using the sixth dataset, which is slightly greater than the transition redshift obtained for interacting THDE model with Hubble horizon and future event horizon as IR cutoff ($z_{t}=0.634^{+0.051}_{-0.045}$ and $z_{t}=0.649^{+0.010}_{-0.025}$ respectively) \cite{sadri2019observational}. The current value of deceleration parameter $(q_{0}=q(z) \rvert_{z=0})$ obtained for the estimated values of the model parameters is $q_{0}=-0.602$ using the fourth dataset and $-0.594$ using the sixth dataset, is closer to the observational value $q_{0}=-0.63\pm0.12$ from $\Lambda$CDM based CMB priors \cite{alam2004case}. Our estimated cosmological parameter values are in good concordance with previous results from model independent methods, parametric reconstruction techniques as well as other dark energy models mentioned in \cite{AlMamon:2018uby,Mamon:2016wow,Mamon:2016dlv,Farooq:2016zwm}.  

\begin{figure}[t!]
	\includegraphics[width=0.45\textwidth]{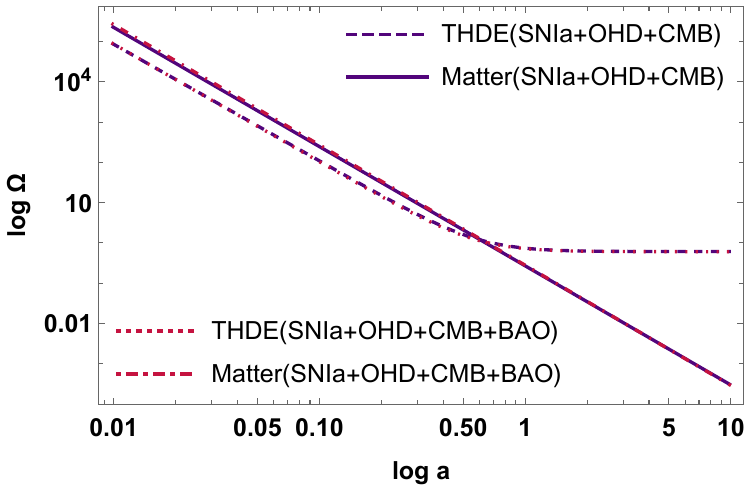}
	\caption{Evolution of matter and THDE density parameters with the scale factor of the universe in logarithmic scale using the fourth and sixth datasets}\label{d_aplot} 
\end{figure}

A juxtaposition of the evolution of matter and THDE density parameters in logarithmic scale is shown in Fig. \ref{d_aplot}. The figure very well implies that THDE will dominate the universe in the future. The evolution of the density parameters of matter and dark energy are comparable with each other for the high redshift values even though the matter density parameter is dominating in the early universe and is diminishing in the late phase of the universe, that solves the coincidence problem as the acceleration began at low redshifts and this is well demonstrated in the Fig. \ref{d_aplot}. The current value of the matter density parameter estimated using the observational constraints is slightly lower than the observational value $\Omega_{mo}=0.3147 \pm 0.0074$ from Planck data \cite{Planck:2018vyg} and the derived value $\Omega_{mo}= 0.29\pm 0.07$ from WMAP results \cite{2003}. Our estimation of the current values of the matter density parameter are similar to the current value obtained for past THDE models \cite{saridakis2018holographic,sadri2019observational} and slightly higher than the value for  the THDE model with future event horizon as IR cutoff \cite{d2019holographic}.

	\subsection{The age of the universe}
The age of the universe can be estimated using the present cosmological observational data, even though systematic and statistical uncertainties will arise during observation and estimation. Theoretically, considering the interacting THDE model, the age of the universe can be calculated by slightly rearranging the equation \eqref{eqn:S1:14} and the resultant equation takes the form
\begin{equation}\label{eqn:S2.3:1}
	H=\frac{\dot{a}}{a}=H_{0}\left(\frac{\Omega_{m0}}{1-b}a^{-3(1-b)}+\dfrac{1}{3}c^\prime a^{-3}+c^{\prime\prime}	\right)^\frac{1}{2}.
\end{equation}
Above equation can be solved by considering the value of scale factor $a=0$ for the big bang time $t_{b}$ and $a=1$ for the present time $t_{0}$ in the following way
\begin{equation*}\label{eqn:S2.3:2}
	\int_{t_b}^{t_0} dt= H_{0}^{-1}\int_{0}^{1}a^{-1} \left(\frac{\Omega_{m0}}{1-b}a^{-3(1-b)}+\dfrac{1}{3}c^\prime a^{-3}+c^{\prime\prime}	\right)^{\frac{-1}{2}}da.
\end{equation*}

Using the best estimated values of the parameters, the age of the universe is evaluated as $14.327$ Gyrs using the fourth dataset and $14.190$ Gyrs using the sixth dataset, which is closer to the age calculated for interacting THDE model with future event horizon as IR cutoff ($14.20^{+0.18}_{-0.32}$~Gyrs) and slightly greater than the age calculated for interacting THDE model with Hubble horizon  as IR cutoff ($13.71^{+0.24}_{-0.41}$~Gyrs, upper bond of $13.43$ and lower bond of $13.04$~Gyrs) \cite{sadri2019observational,ebrahimi2020dynamical}. Our results are closer to the standard value of age $13.8\pm0.02$ Gyrs obtained from Planck mission and $ 13.72 \pm 0.12$ Gyrs from WMAP + BAO + SN data assuming $\Lambda$CDM model \cite{Planck:2018vyg,Hinshaw2008FiveYearWM} and $13.5^{+0.16}_{-0.14}$(stat.) $\pm0.23(0.33)$(sys.) Gyrs from the oldest globular cluster \cite{Valcin:2021jcg}. 
	
\subsection{Evolution in $r-s$ plane and $\omega^{'}_{e}-\omega_{e}$ plane }
	\begin{figure}[t!]
		\includegraphics[width=0.45\textwidth]{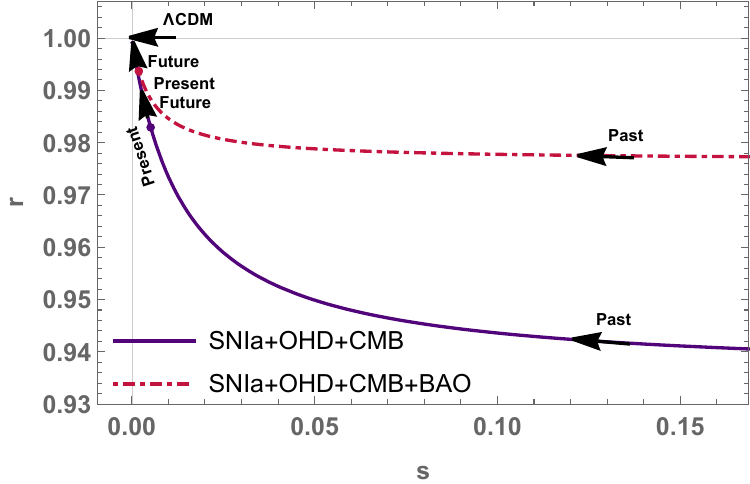} 
		\caption{The statefinder evolutionary trajectory for interacting THDE in the $r-s$ plane for the best estimated values of the model parameters
			using the fourth and sixth datasets}
		\label{r_splot} 
	\end{figure}
To check further, the reliability of THDE model being a generalized model of dark energy in contrast to the present observational data we studied the evolution using the geometrical pair called the statefinder pair $\{r,s\}$ constructed from the scale factor and its derivatives, first defined by Sahni et al. \cite{Sahni:2002fz,Alam:2003sc}. The relationship connecting the statefinder pair and the scale factor of the universe are given by
\begin{equation}\label{eqn:S3.1:1}
	r=\frac{\dddot{a}}{aH^{3}},~~~
	s=\frac{r-1}{3(q-\frac{1}{2})}.
\end{equation}
It is lucid from \eqref{eqn:S3.1:1}, $r$ is third order derivative of `$a$' and $s$ is related to `$r$' and `$q$' linearly. The statefinder pair $\{r,s\}$ can also be expressed in terms of Hubble parameter and its derivatives as
\begin{equation}
	\begin{aligned}\label{eqn:S3.1:2}	
		r & =\frac{1}{2h^{2}}\frac{d^{2}h^{2}}{dx^{2}}+\frac{3}{2h^{2}}\frac{dh^2}{dx}+1 , \\
		s & =-\frac{\frac{1}{2h^{2}}\frac{d^{2}h^{2}}{dx^{2}}+\frac{3}{2h^{2}}\frac{dh^{2}}{dx}}{\frac{3}{2h^{2}}\frac{dh^{2}}{dx}+\frac{9}{2}}.
	\end{aligned}
\end{equation}
Using equation \eqref{eqn:S1:14} in the above equations will results in  the following two expressions in terms of model parameters
\begin{equation}
	\begin{aligned}\label{eqn:S3.1:3}	
		r & =1-\frac{9b\Omega_{m0}a^{-3(1-b)}}{2(	\frac{\Omega_{m0}}{1-b}a^{-3(1-b)}+\frac{1}{3}c^\prime a^{-3}+c^{\prime\prime})} , \\
		s & =\frac{-b\Omega_{m0}a^{-3(1-b)}}{c^{\prime\prime}+\frac{b}{1-b}\Omega_{m0}a^{-3(1-b)}}.
	\end{aligned}
\end{equation}
The  parametric plot of $\{r,s\}$ is obtained as shown in Fig. \ref{r_splot}. Using this diagnostics THDE model can be differentiated from the $\Lambda$CDM
model. The statefinder pair $\{r,s\}$ for $\Lambda$CDM is $\{1,0\}$. Applying the limit $a \to \infty$, shows that the $\{r,s\}$ pair for the model approaches $\{1,0\}$. Thus this model approaches $\Lambda$CDM asymptotically at late times. The present value of $\{r,s\}$ pair from the evolutionary trajectory of the model is obtained as $\{0.983,0.005\}$and $\{0.994,0.002\}$ for the best fit parameter values using the fourth and sixth datasets, respectively. For the quintessence models the $\{r,s\}$ pair lies in $r<1$ and $s>0$. The trajectory in its early phase and in the present shows aforesaid behavior. Comparing the evolutionary trajectories in Fig. \ref{r_splot} for the best fit parameter values from the fourth and sixth datasets tells us that the distance to the $\Lambda$CDM point from the present is larger for the larger values of coupling constant and the distance to the $\Lambda$CDM point from the present is small for the smaller values of coupling constant.

The statefinder evolutionary trajectory of the non interacting THDE model with the Hubble horizon as IR cutoff in the flat FRW universe in the past works \cite{sharma2019diagnosing,Varshney2019fzj,Dubey:2019kzh} shows that for $1<\delta<2$, the model is quintessenc like and approaches $\Lambda$CDM in the future. The interacting THDE model with the Hubble horizon as IR cutoff in flat FRW universe in past works \cite{Varshney2019fzj,Sharma:2019bgp} shows Chaplygin gas like behavior for $1<\delta<2$ and the distance to the $\Lambda$CDM point from the present time will be larger or smaller depending on the corresponding larger or smaller value of coupling constant. In contrast to these results, the evolution of the present model shows a quintessence behavior that finally approaches the $\Lambda$CDM behavior without showing, Chaplygin gas like behavior. In the absence of interaction, the present model shows a behavior almost similar to $\Lambda$CDM. The effect of interaction (i.e., with respect to the value of coupling constant) will only increase or decrease the distance of approach to the $\Lambda$CDM from the present. In general, for the estimated value of $\delta$, the present THDE model shows quintessence behavior and will finally approach $\Lambda$CDM for the form of the interaction considered here (devoid of the coupling constant value) and shows $\Lambda$CDM like behavior in case of zero interaction.		

In addition, we have obtained the effective equation of state parameter using the relation,
\begin{equation}\label{eqn:S2.2:2}
	\omega_{e}=\frac{P}{\rho}=-1-\frac{2\dot{H}}{3H^{2}},
\end{equation}
where $P$ and $\rho$ are the effective pressure and energy density respectively. Using the equation \eqref{eqn:S1:14} in \eqref{eqn:S2.2:2} the effective equation of state parameter for this model takes the following form
\begin{equation}\label{eqn:S2.2:3}
	\omega_{e}=-1+\frac{3\Omega_{m0}e^{-3(1-b)x}+c^\prime e^{-3x}}{\frac{3\Omega_{m0}}{1-b}e^{-3(1-b)x}+c^\prime e^{-3x}+3c^{\prime\prime}}.
\end{equation}
The asymptotic limit of this is as follows. As $x \to -\infty,$ the effective equation of state, $\omega_e \to 0 $ and as $x \to +\infty$, $\omega_e \to -1.$  In the former case, the decelerated epoch, the matter component could be the dominant component, while in the latter case, the accelerated epoch, the dominant component is the dark energy. The derivative of $\omega_{e}$ with respect to $x$ is obtained to be
\begin{equation}
	\omega_{e}^{\prime}=3\left(\frac{x_{3}}{x_{1}}\right)^{2}-\frac{x_{2}}{x_{1}},
\end{equation}
where 
\begin{equation*}
	x_{1}=\frac{3\Omega_{m0}}{1-b}e^{-3(1-b)x}+c^\prime e^{-3x}+3c^{\prime\prime},
\end{equation*}
\begin{equation*}
	x_{2}=9\Omega_{m0}(1-b)e^{-3(1-b)x}+3c^\prime e^{-3x},
\end{equation*}
and
\begin{equation*}
	x_{3}=3\Omega_{m0}e^{-3(1-b)x}+c^\prime e^{-3x}.
\end{equation*}
\begin{figure}[t!]
	\includegraphics[width=0.45\textwidth]{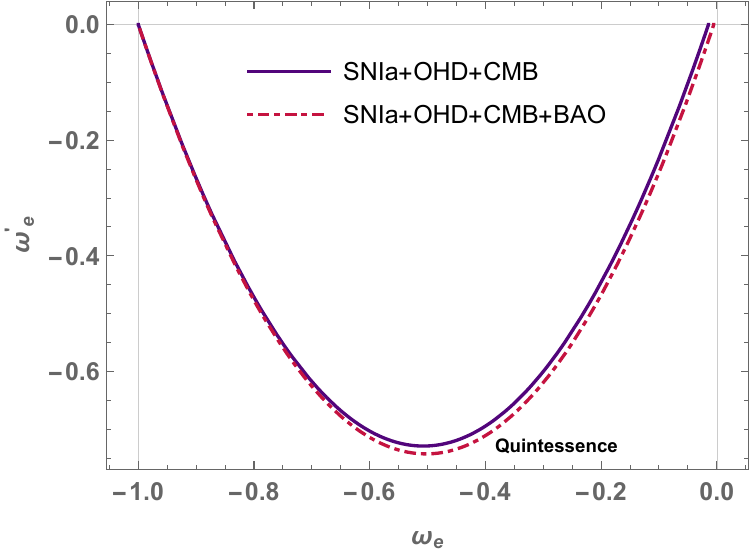} 
	\caption{The evolutionary trajectory for interacting THDE in the $\omega^{\prime}_{e}-\omega_{e}$ plane for the best estimated values of the model parameters using the fourth and sixth datasets}
	\label{we'_weplot} 
\end{figure}
The evolutionary trajectory of the model in the phase plane of $\omega_{e}^{\prime}-\omega_{e}$, the effective equation of state parameter and its derivative, is depicted in Fig. \ref{we'_weplot}.  It complements the results from statefinder analysis by showing that the trajectory is initially confined in the quintessence region $\omega_{e}>-1$, $\omega_{e}^{\prime}<0$  and approaches de Sitter phase in the late time as $\omega_{e} \to-1$, $\omega_{e}^{\prime} \to 0$. The late time cosmic evolution of non interacting THDE model with the Hubble horizon as IR cutoff in flat FRW universe \cite{sharma2019diagnosing,Varshney2019fzj,Dubey:2019kzh} using the $\omega_{de}-\omega^{\prime}_{de}$ plane, where $\omega_{de}$ is the equation of state parameter of dark energy and $\omega^{\prime}_{de}$ is its evolution, shows that for $1<\delta<2$, the model shows quintessence behavior. In the presence of  interaction \cite{Varshney2019fzj,Sharma:2019bgp}, the model shows completely phantom behavior with higher coupling constant values. Whereas the present THDE model shows quintessence behavior in general and approaches de Sitter phase in the late time in the $\omega^\prime_{e}-\omega_{e}$ plane as well.
\section{Dynamical analysis of interacting THDE model}
\label{sec:3}
The dynamical system analysis enables us to see whether the model shows the stable evolution consistent with the observations. For the phase space analysis of the model, a set of dimensionless variables is introduced as follows:
\begin{equation}\label{eqn:S3.3:1}
	u=\frac{\rho_{m}}{3H^{2}},~~~ v=\frac{\rho_{de}}{3H^{2}}.
\end{equation}
The autonomous equations in terms of phase space variables can be obtained using the Friedmann equation as,
\begin{equation}\label{eqn:S3.3:2}
	u^\prime =u^{2}-2uv+(3b-1)u,
\end{equation}
\begin{equation}\label{eqn:S3.3:3}
	\begin{split}
		v^\prime=&\frac{(2-\delta)[2(\alpha^\prime-\beta^\prime)(v^{2}-v)-uv((\alpha^\prime-\beta^\prime)+\frac{3\beta^\prime}{2}(3b-1))]}{\alpha^\prime+\beta^\prime(v-\frac{u}{2}-1)}\\
		&-2v^{2}+uv+2v,
	\end{split}
\end{equation}
where prime $(^\prime)$ denotes differentiation with respect to $x=\ln a.$ Both the equations are functions of $u$ and $v$. By equating $u^\prime=0$ and $v^\prime=0$, the real and physically meaningful critical points obtained are $(\tilde{u},\tilde{v})= (0,1),(1-3b, 0)$ and they corresponds to the de Sitter phase and the matter dominated phase respectively.

The behavior of the trajectory near the critical points can be analyzed by considering the small neighbourhood of critical points as
\begin{equation}\label{eqn:S3.3:4}
	u=\tilde{u}+\xi,\quad
	v=\tilde{v}+\eta.
\end{equation}
in which $\xi$ and $\eta$ are small compared to $\tilde{u}$ and $\tilde{v}$. Linearizing the set of equations \eqref{eqn:S3.3:2} and \eqref{eqn:S3.3:3} with respect to $\xi$ and $\eta$, will result in a matrix equation
\begin{equation}\label{eqn:S3.3:5}
	\begin{bmatrix}
		\xi^\prime\\ 
		\eta^\prime	
	\end{bmatrix}= \begin{bmatrix}
		\left(\frac{\partial u^\prime}{\partial u}\right)_{\left(\tilde{u},\tilde{v}\right)} & \left(\frac{\partial u^\prime}{\partial v}\right)_{\left(\tilde{u},\tilde{v}\right)} \\
		\left(\frac{\partial v^\prime}{\partial u}\right)_{\left(\tilde{u},\tilde{v}\right)} &
		\left(\frac{\partial v^\prime}{\partial v} \right)_{\left(\tilde{u},\tilde{v}\right)}
	\end{bmatrix} 
	\begin{bmatrix}
		\xi \\
		\eta
	\end{bmatrix},
\end{equation}
where $2 \times 2$ matrix on the right hand side of the above equation is the Jacobian at the critical points corresponding to the autonomous system and the partial derivatives are calculated about the critical points $(\tilde{u},\tilde{v})$. 

\begin{figure}[t!]
	\centering
	\includegraphics[width=0.45\textwidth]{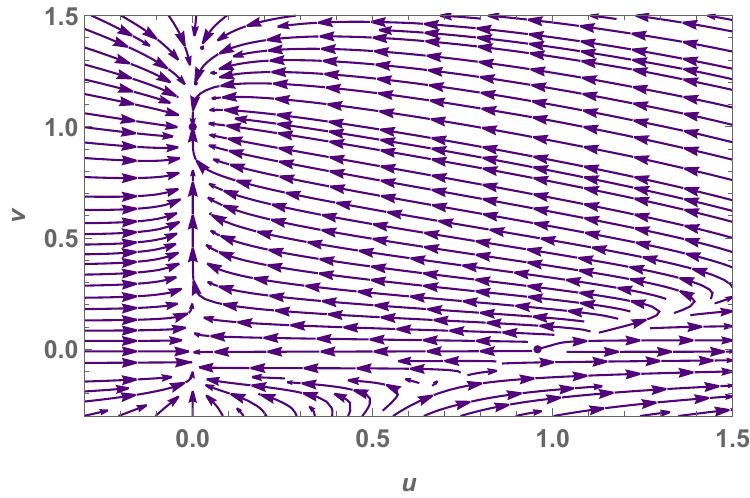}   
	\caption{The evolution of the phase space trajectories in the $u-v$ plane for the best estimated model parameters using SNIa$+$OHD$+$CMB dataset} 
	\label{psaplot} 
\end{figure}

\begin{figure}[t!]
	\centering
	\includegraphics[width=0.45\textwidth]{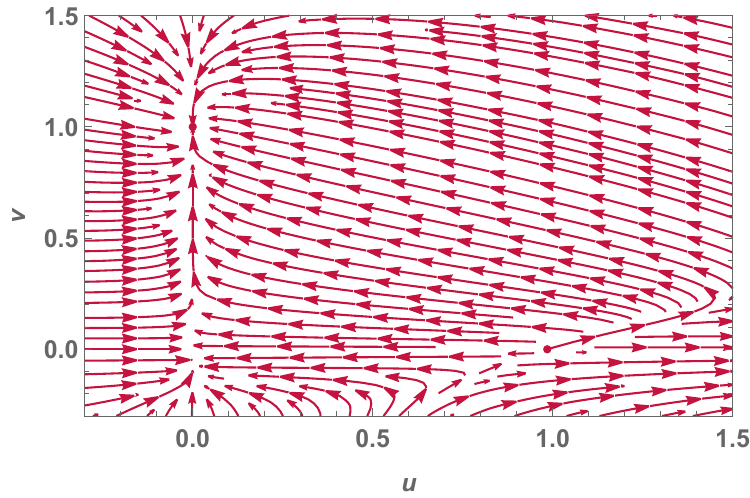}   
	\caption{The evolution of the phase space trajectories in the $u-v$ plane for the best estimated model parameters using SNIa$+$OHD$+$CMB$+$BAO dataset}
	\label{psaplot2} 
\end{figure}

The eigenvalues of the $2 \times 2$ Jacobian matrix can be obtained by diagonalizing the matrix, and its nature decides the asymptotic stability nature of critical points. The critical points (equilibrium points) can be source point or past attractor if the eigenvalues were positive values, or a saddle point if at least one of the eigenvalues were of a different sign, and future attractor if all the eigenvalues were negative values \cite{gregory_2006}. 
\begin{table}[b]
	\caption{Critical points and stability of the corresponding eigenvalues calculated using the best estimated model parameter values}
	\label{tab:2} 
	\begin{tabular}{llll}
		\hline\noalign{\smallskip}
		Data & Critical points & Eigenvalues &Nature\\ 
		\noalign{\smallskip}\hline\noalign{\smallskip}
		SNIa$+$OHD$+$ & $(0.958, 0)$ & $(0.958,0.045)$ & Unstable\\
		CMB& $(0,1)$ & $(-2.958, -0.559)$ & Stable\\
		\noalign{\smallskip}\hline\noalign{\smallskip}
		SNIa$+$OHD$+$ & $(0.985, 0)$ & $(0.985, 0.054)$ & Unstable\\
		CMB$+$BAO & $(0,1)$ & $(-2.985, -0.657)$ & Stable\\
		\noalign{\smallskip}\hline	
	\end{tabular}
\end{table}
The eigenvalues corresponding to the Jacobian matrix are evaluated and are given in Table \ref{tab:2}. The critical points $(0.958,0)$ and $(0.985,0)$ for which the matter is the dominant component, are unstable since their eigenvalues are positive. All trajectories diverge from $(0.958,0)$ and $(0.985,0)$, which is clear from Figs. \ref{psaplot} and \ref{psaplot2}. The critical point $(0,1),$ for which dark energy is the dominant component, is stable since its eigenvalues are negative. This represents a future stable point. The convergence of all trajectories to the future attractor at $(0,1)$ is well depicted in the phase space diagram shown in Figs. \ref{psaplot} and \ref{psaplot2}. Universe undergoes a de Sitter expansion at this point. Our analysis shows that the system emerges from a decelerated expansion and ends on a de Sitter epoch. The previous works \cite{huang2019stability,ebrahimi2020dynamical} in the dynamical system analysis of interacting and non interacting THDE with Hubble horizon as IR cutoff also show a stable point corresponding to the de Sitter phase in the future.

	\section{Thermodynamics of interacting THDE model}
\label{sec:4}
Any isolated macroscopic system advances to the maximum entropy state in compliance with the constraints of the generalized second law (GSL) of thermodynamics \cite{Pavon:2012qn}. From the aforesaid statement, it can be deduced that the entropy $S$, of isolated systems, cannot decrease, i.e., $S^\prime \geq 0$, where the prime means derivative with respect to the scale factor $a$ and it must be a convex function of the scale factor, $S''<0$, at least at the last juncture of the evolution so that the system attains a stable equilibrium state. 
\begin{figure}[t!]
	\includegraphics[width=0.45\textwidth]{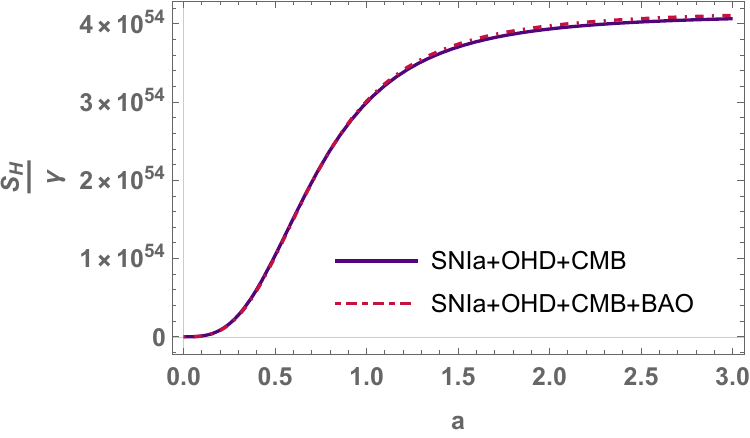} 
	\caption{The evolution of $\frac{S_{H}}{\gamma}$ with scale factor for the best estimated values of the model parameters using the fourth and sixth datasets}
	\label{s_aplot} 
\end{figure}

In the present work, we consider Tsallis entropy, given in equation \eqref{eqn:S1:1}  as the horizon entropy in place of the standard Bekenstein-Hawking entropy. 
Then the horizon entropy of the apparent horizon of a flat universe, $S_H$ becomes
\begin{equation}\label{eqn:S3.2:2}
	S_{H}=\gamma \left( \frac{4\pi c^{2}}{H^{2}}\right)^{\delta} k_{B},
\end{equation}
where $c$ is the speed of light and $k_{B}$ is the Boltzmann constant. The evolution of $\frac{S_{H}}{\gamma}$ with respect to the scale factor is shown in Fig. \ref{s_aplot}. The rate of change of  horizon entropy with respect to scale factor is given by
\begin{equation}\label{eqn:S3.2:3}
	S_{H}^\prime=\gamma \left( 4\pi c^{2}\right)^{\delta}\left(\frac{-2\delta H^\prime}{H^{{2\delta+1}}}\right).
\end{equation}
\begin{figure}[t!]
	\includegraphics[width=0.45\textwidth]{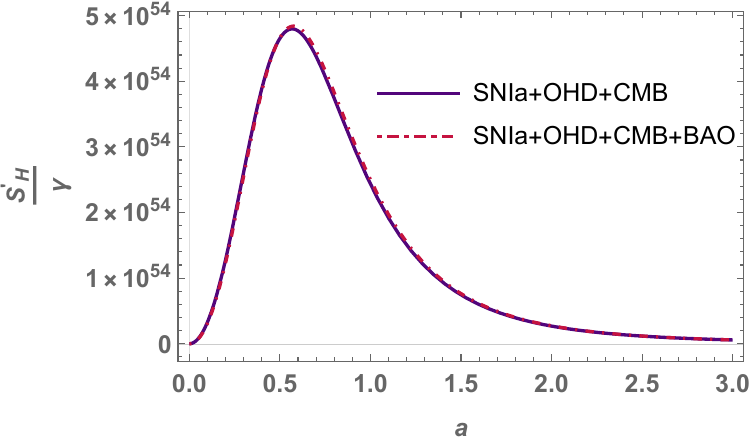} 
	\caption{The evolution of $\frac{S_{H}^\prime}{\gamma}$ with scale factor for the best estimated values of the model parameters using the fourth and sixth datasets}
	\label{s'_aplot} 
\end{figure}
The second derivative of horizon entropy with respect to the scale factor is given by
\begin{equation}\label{eqn:S3.2:4}
	S_{H}^{\prime\prime}=2\gamma \delta\left( 4\pi c^{2}\right)^{\delta}\left[(2\delta+1)\left(\frac{H^\prime}{H^{\delta+1}}\right)^2-\frac{ H^{\prime\prime}}{H^{{2\delta+1}}} \right].
\end{equation}	
\begin{figure}[t!]
	\centering
	\includegraphics[width=\columnwidth]{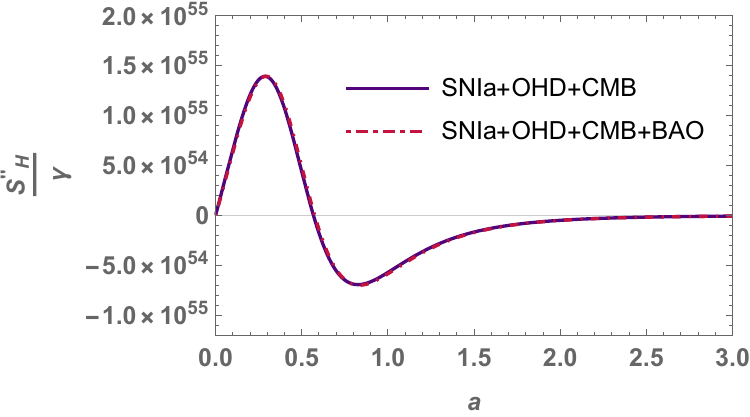} 
	\caption{The evolution of $\frac{S_{H}^{\prime\prime}}{\gamma}$ with scale factor for the best estimated values of the model parameters using the fourth and sixth datasets}\label{s"_aplot} 	
\end{figure}

The numerical simulations  and analysis of the observational data on Hubble parameter shows that $H^\prime<0$ and $H^{\prime\prime} >0$ \cite{Crawford:2010rg, Carvalho:2011qw,Simon:2004tf, Stern:2009ep}. Fig. \ref{s'_aplot} clearly shows that the horizon entropy satisfies  $\frac{S_{H}^\prime}{\gamma}\geq 0$, always, which in turn implies that the horizon entropy is always increasing. Finally, in the asymptotic limit $a \to \infty$, the entropy $S_{H}^{\prime}$ approaches zero, consequently horizon entropy attains a constant value. That is, the horizon entropy asymptotically reaches  constant at the end de Sitter epoch. The behavior of $\frac{S_{H}^{\prime\prime}}{\gamma}$ with scale factor in Fig.~\ref{s"_aplot} guarantees the convexity, thereby ensuring that entropy does not grow unboundedly. 

The GSL of thermodynamics stipulates that the entropy of the horizon together with the entropy of the matter inside the horizon must always increase with time \cite{PhysRevD.74.083520}. Since it is clear the matter entropy is much less than the horizon entropy (smaller in the order of 35) \cite{egan2010larger,B:2020wuq}, the total entropy of the universe can approximately be taken as the horizon entropy \cite{PhysRevD.15.2738}. Since the horizon entropy is an increasing function as shown in Fig. \ref{s_aplot} and Fig. \ref{s'_aplot}, the GSL is considered to be satisfied. 
Certain past studies \cite{al2020generalized} of interacting THDE model with Hubble horizon as IR cutoff with the form of interaction $Q=3H(b^{2}_{1} \rho_{m}+b^{2}_{2} \rho_{de})$, where $b^{2}_{1,2}$ are the coupling constants, and with the varying equation of state of DE, shows the plausibility of violation of GSL depending on the evolution of the universe. Unlike those studies, the present analysis by considering DE as dynamical vacuum guarantees the validity of GSL throughout the evolution of the universe. 

As evident from the equation \eqref{hbzero}, when the interaction parameter $b=0$, the present model reduces to a model, like $\Lambda$CDM with an effective cosmological constant $\tilde{\Omega}_{de0},$ and the value of which is around $0.735$ and $0.729$ for the best estimated parameter values using the fourth and sixth datasets, respectively. Results from various studies, like in \cite{2021EPJC...81..644S,2013EPJC...73.2619M}, support that a $\Lambda$CDM like model would satisfy the GSL. Following this, it can be concluded that the present model with $b=0$ will satisfy the GSL throughout the evolution.

\section{Conclusion} \label{6}
Recently much interest has been arisen in the new holographic dark energy based on the non extensive entropy, Tsallis entropy, and holographic principle. In the present work, we have analyzed interacting Tsallis Holographic Dark Energy (THDE) as dynamical vacuum to explain the recent accelerated cosmic expansion. We have considered the Granda-Oliveros (GO) scale, a function of Hubble parameter and its time derivative, as IR cutoff. The interaction between the dark sectors has been accounted, by a simple form of interaction term $Q=3bH\rho_{m}$.

An exact solution for the Hubble parameter in terms of scale factor was obtained by analytically solving the Friedmann equation with the energy conservation equation. The solution successfully explains the prior decelerated expansion and the late time accelerated expansion, hence effectively elucidating the transition. We have imposed the observational constraints on the cosmological parameters of the interacting THDE model using the Pantheon sample, the OHD, the CMB and the BAO data. The estimated current values of the Hubble parameter ($H_{0}=68.867^{+1.199}_{-1.159}$  kms$^{-1}$Mpc$^{-1}$ from the fourth dataset and $H_{0}=68.672^{+1.196}_{-1.141}$  kms$^{-1}$Mpc$^{-1}$ from  the sixth dataset) and the matter density parameter ($\Omega_{m0}=0.271^{+0.022}_{-0.015}$ from the fourth dataset and $\Omega_{m0}=0.281^{+0.017}_{-0.015}$ from the sixth dataset) are consistent with the observational results. The value of the coupling constant is positive, and hence it obeys the Le Chatelier-Braun principle. 

The cosmological evolution of the deceleration parameter shows that the end phase will be a de Sitter one, $q \to -1$ as $z \to -1$ and the transition from prior matter dominated epoch to the late THDE dominated epoch is found to occur at a redshift around $z_{t} \sim 0.8$. The studies on the expansion profile of the THDE density and the matter density explains the dominating and diminishing nature of the matter density in the early phase and late phase, respectively, and along with that, similar nature of the evolution of THDE in the early phase removes the coincidence problem. The universe's age that can be intuited from the interacting THDE model is around 14 Gyrs from the best estimated values of model parameters. Recent observations do subsistence the results. 

The diagnosis of the interacting THDE model is carried out using the statefinder analysis and the $\omega^\prime_e-\omega_e$ pair. The evolutionary trajectory of the interacting THDE model shows that the model is distinguishably different from the standard $\Lambda$CDM and shows a quintessence behavior in the early phase and approaches $\Lambda$CDM in the far future. The trajectories of the $\omega^\prime_e-\omega_e$  affirm the results from the statefinder diagnostics. The critical points obtained from the phase space analysis of the model shows that the de Sitter phase is a stable equilibrium, and matter dominated phase is an unstable equilibrium. Thus the dynamical system analysis of the interacting THDE model portrays a consistent background evolution of the universe from the prior decelerated phase to the late accelerated phase. 

The thermodynamical analysis of the THDE model sho-ws that the horizon entropy increases with cosmic time and achieve the maximization condition during the late epoch. It can be concluded that the generalized second law of thermodynamics remains valid in the dynamical vacuum treatment of the model by considering the dominant contribution of the horizon entropy to the total entropy of the universe. It is to be noted that, for zero interaction parameter, the model will reduce to a $\Lambda$CDM like one, where the validity of the laws of thermodynamics is guaranteed.
\begin{acknowledgments}
We are thankful to the referee for the valuable and insightful comments which helped in bringing improvements in the current manuscript. We are thankful to Dr. Jerin Mohan N D and Hassan Basari V T for their valuable suggestions in data analysis. We are thankful to Ms. Rosemin John for pointing out the error in Table \ref{tab:1} in the published version. M Dheepika is thankful to CSIR for the financial support through JRF.
\end{acknowledgments}


\begin{thebibliography}{200}
	\bibitem{SupernovaSearchTeam:1998fmf}
	A.G. Riess, et~al., Astron. J. \textbf{116}, 1009 (1998)
	
	\bibitem{SupernovaCosmologyProject:1998vns}
	S.~Perlmutter, et~al., Astrophys. J. \textbf{517}, 565 (1999)
	
	\bibitem{Caldwell:1997ii}
	R.R. Caldwell, R.~Dave, P.J. Steinhardt, Phys. Rev. Lett. \textbf{80}, 1582
	(1998)
	
	\bibitem{zlatev1999quintessence}
	I.~Zlatev, L.~Wang, P.J. Steinhardt, Phys. Rev. Lett. \textbf{82}(5), 896
	(1999)
	
	\bibitem{Armendariz-Picon:2000nqq}
	C.~Armendariz-Picon, V.F. Mukhanov, P.J. Steinhardt, Phys. Rev. Lett.
	\textbf{85}, 4438 (2000)
	
	\bibitem{Caldwell:1999ew}
	R.R. Caldwell, Phys. Lett. B \textbf{545}, 23 (2002)
	
	\bibitem{tHooft:1993dmi}
	G.~'t~Hooft, Conf. Proc. C \textbf{930308}, 284 (1993)
	
	\bibitem{Susskind:1994vu}
	L.~Susskind, J. Math. Phys. \textbf{36}, 6377 (1995)
	
	\bibitem{PhysRevLett.82.4971}
	A.G. Cohen, D.B. Kaplan, A.E. Nelson, Phys. Rev. Lett. \textbf{82}, 4971 (1999)
	
	\bibitem{2005PhLB..610...18M}
	Y.S. {Myung}, Phys. Lett. B \textbf{610}(1-2), 18 (2005)
	
	\bibitem{pavon2007holographic}
	D.~Pav{\'o}n, J. Phys. A Math. Theor. \textbf{40}(25), 6865 (2007)
	
	\bibitem{Granda:2008dk}
	L.N. Granda, A.~Oliveros, Phys. Lett. B \textbf{669}, 275 (2008)
	
	\bibitem{fischler1998holography}
	W.~Fischler, L.~Susskind, arXiv preprint hep-th/9806039  (1998)
	
	\bibitem{li2004model}
	M.~Li, Phys. Lett. B \textbf{603}(1-2), 1 (2004)
	
	\bibitem{wetterich1994cosmon}
	C.~Wetterich, Astron. Astrophys. \textbf{301}, 321 (1995)
	
	\bibitem{bertolami2007dark}
	O.~Bertolami, F.G. Pedro, M.~Le~Delliou, Phys. Lett. B \textbf{654}(5-6), 165
	(2007)
	
	\bibitem{olivares2005observational}
	G.~Olivares, F.~Atrio-Barandela, D.~Pavon, Phys. Rev. D \textbf{71}(6), 063523
	(2005)
	
	\bibitem{pavon2005holographic}
	D.~Pavon, W.~Zimdahl, Phys. Lett. B \textbf{628}(3-4), 206 (2005)
	
	\bibitem{zimdahl2007interacting}
	W.~Zimdahl, D.~Pav{\'o}n, Class. Quant. Grav. \textbf{24}(22), 5461 (2007)
	
	\bibitem{2009PhRvD..79d3511G}
	C.~{Gao}, F.~{Wu}, X.~{Chen}, Y.G. {Shen}, Phys. Rev. D \textbf{79}(4), 043511
	(2009)
	
	\bibitem{Nojiri:2005pu}
	S.~Nojiri, S.D. Odintsov, Gen. Rel. Grav. \textbf{38}, 1285 (2006)
	
	\bibitem{PhysRevD.84.123507}
	L.P. Chimento, M.G. Richarte, Phys. Rev. D \textbf{84}, 123507 (2011)
	
	\bibitem{2012PhRvD..85l7301C}
	L.P. {Chimento}, M.G. {Richarte}, Phys. Rev. D \textbf{85}(12), 127301 (2012)
	
	\bibitem{2013EPJC...73.2352C}
	L.P. {Chimento}, M.G. {Richarte}, Eur. Phys. J. C \textbf{73}, 2352 (2013)
	
	\bibitem{2017EPJC...77..528N}
	S.~{Nojiri}, S.D. {Odintsov}, Eur. Phys. J. C \textbf{77}(8), 528 (2017)
	
	\bibitem{das2002general}
	S.~Das, P.~Majumdar, R.K. Bhaduri, Class. Quant. Grav. \textbf{19}(9), 2355
	(2002)
	
	\bibitem{das2008black}
	S.~Das, S.~Shankaranarayanan, S.~Sur, \emph{{Black hole entropy from
			entanglement: a review}}, vol. 268 (2009)
	
	\bibitem{das2008power}
	S.~Das, S.~Shankaranarayanan, S.~Sur, Phys. Rev. D \textbf{77}(6), 064013
	(2008)
	
	\bibitem{Das:2010su}
	S.~Das, S.~Shankaranarayanan, S.~Sur, in \emph{{12th Marcel Grossmann Meeting
			on General Relativity}} (2010), pp. 1138--1141
	
	\bibitem{radicella2010generalized}
	N.~Radicella, D.~Pav{\'o}n, Phys. Lett. B \textbf{691}(3), 121 (2010)
	
	\bibitem{das2012entanglement}
	S.~Das, S.~Shankaranarayanan, S.~Sur, in \emph{The Twelfth Marcel Grossmann
		Meeting: On Recent Developments in Theoretical and Experimental General
		Relativity, Astrophysics and Relativistic Field Theories (In 3 Volumes)}
	(World Scientific, 2012), pp. 1138--1141
	
	\bibitem{Tsallis:1987eu}
	C.~Tsallis, J. Statist. Phys. \textbf{52}, 479 (1988)
	
	\bibitem{Lyra:1997ggy}
	M.L. Lyra, C.~Tsallis, Phys. Rev. Lett. \textbf{80}, 53 (1998)
	
	\bibitem{tsallis1998role}
	C.~Tsallis, R.~Mendes, A.R. Plastino, Phys. A: Stat. Mech. Appl.
	\textbf{261}(3-4), 534 (1998)
	
	\bibitem{wilk2000interpretation}
	G.~Wilk, Z.~W{\l}odarczyk, Phys. Rev. Lett. \textbf{84}(13), 2770 (2000)
	
	\bibitem{Tsallis:2012js}
	C.~Tsallis, L.J.L. Cirto, Eur. Phys. J. C \textbf{73}, 2487 (2013)
	
	\bibitem{2014arXiv1403.5706N}
	R.d.C. {Nunes}, J.~{Barboza}, Ed{\'e}sio~M., E.M.C. {Abreu}, J.~{Ananias Neto},
	arXiv e-prints arXiv:1403.5706 (2014)
	
	\bibitem{2016JCAP...08..051N}
	R.C. {Nunes}, J.~{Barboza}, Ed{\'e}sio~M., E.M.C. {Abreu}, J.~{Ananias Neto},
	JCAP \textbf{2016}(8), 051 (2016)
	
	\bibitem{moradpour2016implications}
	H.~Moradpour, Int. J. Theor. Phys. \textbf{55}(9), 4176 (2016)
	
	\bibitem{tavayef2018tsallis}
	M.~Tavayef, A.~Sheykhi, K.~Bamba, H.~Moradpour, Phys. Lett. B \textbf{781}, 195
	(2018)
	
	\bibitem{zadeh2018note}
	M.A. Zadeh, A.~Sheykhi, H.~Moradpour, K.~Bamba, Eur. Phys. J. C
	\textbf{78}(11), 1 (2018)
	
	\bibitem{Aly2019TsallisHD}
	A.A. Aly, Adv. Astron. \textbf{2019}, 8138067 (2019)
	
	\bibitem{Srivastava:2020hng}
	V.~Srivastava, U.K. Sharma, Int. J. Geom. Meth. Mod. Phys. \textbf{17}(11),
	2050144 (2020)
	
	\bibitem{Sharma:2021dqj}
	U.K. Sharma, V.~Srivastava, New Astron. \textbf{84}, 101519 (2021)
	
	\bibitem{AbdollahiZadeh2019ThermalSO}
	M.A. Zadeh, A.~Sheykhi, H.~Moradpour, Gen. Relativ. Gravit. \textbf{51}, 1
	(2019)
	
	\bibitem{DIXIT2019101281}
	A.~Dixit, U.K. Sharma, A.~Pradhan, New Astron. \textbf{73}, 101281 (2019)
	
	\bibitem{aditya2019observational}
	Y.~Aditya, S.~Mandal, P.~Sahoo, D.~Reddy, Eur. Phys. J. C \textbf{79}(12), 1
	(2019)
	
	\bibitem{Yadav:2020wsd}
	A.K. Yadav, Eur. Phys. J. C \textbf{81}(1), 8 (2021)
	
	\bibitem{Sharma:2019bgp}
	U.K. Sharma, V.C. Dubey, A.~Pradhan, Int. J. Geom. Meth. Mod. Phys.
	\textbf{17}(02), 2050032 (2020)
	
	\bibitem{Saha:2020vxn}
	A.~Saha, S.~Ghose, Astrophys. Space Sci. \textbf{365}(6), 98 (2020)
	
	\bibitem{ghaffari2020holographic}
	S.~Ghaffari, A.~Mamon, H.~Moradpour, A.~Ziaie, Mod. Phys. Lett. A
	\textbf{35}(33), 2050276 (2020)
	
	\bibitem{astashenok2020some}
	A.V. Astashenok, A.S. Tepliakov, Int. J. Mod. Phys. D \textbf{29}(01), 1950176
	(2020)
	
	\bibitem{Ghaffari:2019qcv}
	S.~Ghaffari, E.~Sadri, A.H. Ziaie, Mod. Phys. Lett. A \textbf{35}(14), 2050107
	(2020)
	
	\bibitem{al2020study}
	A.~Al~Mamon, Mod. Phys. Lett. A \textbf{35}(30), 2050251 (2020)
	
	\bibitem{al2020generalized}
	A.~Al~Mamon, A.H. Ziaie, K.~Bamba, Eur. Phys. J. C \textbf{80}(10), 1 (2020)
	
	\bibitem{jawad2021cosmic}
	A.~Jawad, A.M. Sultan, Adv. High Energy Phys. \textbf{2021} (2021)
	
	\bibitem{Nojiri:2021iko}
	S.~Nojiri, S.D. Odintsov, T.~Paul, Symmetry \textbf{13}
	
	\bibitem{dubey2019tsallis}
	V.C. Dubey, S.~Srivastava, U.K. Sharma, A.~Pradhan, Pramana \textbf{93}(5), 1
	(2019)
	
	\bibitem{ChandraDubey:2020tng}
	V.~Chandra~Dubey, A.~Kumar~Mishra, S.~Srivastava, U.~Kumar~Sharma, Int. J.
	Geom. Meth. Mod. Phys. \textbf{17}(01), 2050011 (2020)
	
	\bibitem{AbdollahiZadeh:2019cqi}
	M.~Abdollahi~Zadeh, A.~Sheykhi, K.~Bamba, H.~Moradpour, Mod. Phys. Lett. A
	\textbf{35}(09), 2050053 (2019)
	
	\bibitem{sharma2020swampland}
	U.K. Sharma, S.~Srivastava, A.~Beesham, Int. J. Geom. Meth. Mod. Phys.
	\textbf{17}(07), 2050098 (2020)
	
	\bibitem{Korunur:2019rhg}
	M.~Korunur, Mod. Phys. Lett. A \textbf{34}(37), 1950310 (2019)
	
	\bibitem{Bhattacharjee_2020}
	S.~Bhattacharjee, Astrophys. Space Sci. \textbf{365}(6) (2020)
	
	\bibitem{huang2019stability}
	Q.~Huang, H.~Huang, J.~Chen, L.~Zhang, F.~Tu, Class. Quant. Grav.
	\textbf{36}(17), 175001 (2019)
	
	\bibitem{Varshney2019fzj}
	G.~Varshney, U.K. Sharma, A.~Pradhan, New Astron. \textbf{70}, 36 (2019)
	
	\bibitem{sharma2019diagnosing}
	U.K. Sharma, A.~Pradhan, Mod. Phys. Lett. A \textbf{34}(13), 1950101 (2019)
	
	\bibitem{jawad2019non}
	A.~Jawad, S.~Rani, N.~Azhar, Mod. Phys. Lett. A \textbf{34}(07n08), 1950055
	(2019)
	
	\bibitem{zhang2020diagnosing}
	N.~Zhang, Y.B. Wu, J.N. Chi, Z.~Yu, D.F. Xu, Mod. Phys. Lett. A
	\textbf{35}(08), 2050044 (2020)
	
	\bibitem{srivastava2020statefinder}
	V.~Srivastava, U.K. Sharma, New Astron. \textbf{78}, 101380 (2020)
	
	\bibitem{IQBAL2019100349}
	A.~Iqbal, A.~Jawad, Phys. Dark Univ. \textbf{26}, 100349 (2019)
	
	\bibitem{mohammadi2021tsallis}
	A.~Mohammadi, T.~Golanbari, K.~Bamba, I.P. Lobo, Phys. Rev. D \textbf{103}(8),
	083505 (2021)
	
	\bibitem{dubey2021comparing}
	V.C. Dubey, U.K. Sharma, New Astron. \textbf{86}, 101586 (2021)
	
	\bibitem{d2019holographic}
	R.~D’Agostino, Phys. Rev. D \textbf{99}(10), 103524 (2019)
	
	\bibitem{younas2019cosmological}
	M.~Younas, A.~Jawad, S.~Qummer, H.~Moradpour, S.~Rani, Adv. High Energy Phys.
	\textbf{2019} (2019)
	
	\bibitem{da2021cosmological}
	W.~da~Silva, R.~Silva, Eur. Phys. J. Plus \textbf{136}(5), 1 (2021)
	
	\bibitem{aly2019study}
	A.A. Aly, Eur. Phys. J. Plus \textbf{134}(7), 1 (2019)
	
	\bibitem{sym11010092}
	M.~Sharif, S.~Saba, Symmetry \textbf{11}(1) (2019)
	
	\bibitem{Jawad:2019ouc}
	A.~Jawad, A.~Aslam, S.~Rani, Int. J. Mod. Phys. D \textbf{28}(11), 1950146
	(2019)
	
	\bibitem{shaikh2021diagnosing}
	A.~Shaikh, arXiv:2105.04411  (2021)
	
	\bibitem{ens2020f}
	P.~Ens, A.~Santos, EPL \textbf{131}(4), 40007 (2020)
	
	\bibitem{jawad2020generalized}
	A.~Jawad, S.~Hussain, S.~Rani, S.~Qummer, Int. J. Geom. Methods Mod. Phys.
	\textbf{17}(08), 2050124 (2020)
	
	\bibitem{jawad2020generalizedT}
	A.~Jawad, S.~Hussain, Int. J. Geom. Meth. Mod. Phys.  (2020)
	
	\bibitem{santhi2020bianchi}
	M.V. Santhi, Y.~Sobhanbabu, Eur. Phys. J. C \textbf{80}(12), 1 (2020)
	
	\bibitem{aly2020cosmological}
	A.A. Aly, M.A. Elrashied, M.M. Selim, Int. J. Mod. Phys. D \textbf{29}(03),
	2050023 (2020)
	
	\bibitem{rani2019cosmological}
	S.~Rani, A.~Jawad, K.~Bamba, I.U. Malik, Symmetry \textbf{11}(4), 509 (2019)
	
	\bibitem{Shekh:2021bgh}
	S.H. Shekh, P.H.R.S. Moraes, P.K. Sahoo, Universe \textbf{7}(3), 67 (2021)
	
	\bibitem{Varshney:2020eun}
	G.~Varshney, U.K. Sharma, A.~Pradhan, Eur. Phys. J. Plus \textbf{135}(7), 541
	(2020)
	
	\bibitem{sharma2020reconstruction}
	U.K. Sharma, arXiv preprint arXiv:2005.03979  (2020)
	
	\bibitem{Varshney:2021xvg}
	G.~Varshney, U.K. Sharma, A.~Pradhan, N.~Kumar, Chin. J. Phys. \textbf{73},
	1474 (2021)
	
	\bibitem{VIJAYASANTHI2021101648}
	M.~{Vijaya Santhi}, Y.~Sobhanbabu, New Astron. \textbf{89}, 101648 (2021)
	
	\bibitem{maity2020study}
	S.~Maity, U.~Debnath, Int. J. Geom. Meth. Mod. Phys. \textbf{17}(11), 2050170
	(2020)
	
	\bibitem{Liu:2021heo}
	Y.~Liu, Eur. Phys. J. Plus \textbf{136}(5), 579 (2021)
	
	\bibitem{Zubair:2021yrq}
	M.~Zubair, L.R. Durrani, Chin. J. Phys. \textbf{69}, 153 (2021)
	
	\bibitem{bhattacharjee2021growth}
	S.~Bhattacharjee, Eur. Phys. J. C \textbf{81}(3), 1 (2021)
	
	\bibitem{sym10110635}
	A.~Jawad, K.~Bamba, M.~Younas, S.~Qummer, S.~Rani, Symmetry \textbf{10}(11)
	(2018)
	
	\bibitem{pradhan2021tsallis}
	A.~Pradhan, A.~Dixit, New Astron. p. 101636 (2021)
	
	\bibitem{lymperis2018modified}
	A.~Lymperis, E.N. Saridakis, Eur. Phys. J. C \textbf{78}(12), 1 (2018)
	
	\bibitem{sheykhi2018modified}
	A.~Sheykhi, Phys. Lett. B \textbf{785}, 118 (2018)
	
	\bibitem{nojiri2019modified}
	S.~Nojiri, S.D. Odintsov, E.N. Saridakis, Eur. Phys. J. C \textbf{79}(3), 1
	(2019)
	
	\bibitem{nojiri2020correspondence}
	S.~Nojiri, S.D. Odintsov, E.N. Saridakis, R.~Myrzakulov, Nucl. Phys. B.
	\textbf{950}, 114850 (2020)
	
	\bibitem{abbasi2020tsallisian}
	K.~Abbasi, S.~Gharaati, arXiv:2006.01763  (2020)
	
	\bibitem{asghari2021observational}
	M.~Asghari, A.~Sheykhi, arXiv:2106.15551  (2021)
	
	\bibitem{George:2018myt}
	P.~George, V.M. Shareef, T.K. Mathew, Int. J. Mod. Phys. D \textbf{28}(04),
	1950060 (2018)
	
	\bibitem{Starobinsky:1998fr}
	A.A. Starobinsky, JETP Lett. \textbf{68}, 757 (1998)
	
	\bibitem{saridakis2018holographic}
	E.N. Saridakis, K.~Bamba, R.~Myrzakulov, F.K. Anagnostopoulos, JCAP
	\textbf{2018}(12), 012 (2018)
	
	\bibitem{Dubey:2019kzh}
	V.C. Dubey, U.K. Sharma, A.~Beesham, Int. J. Mod. Phys. D \textbf{28}(15),
	1950164 (2019)
	
	\bibitem{ebrahimi2020dynamical}
	E.~Ebrahimi, Astrophys. Space Sci. \textbf{365}, 1 (2020)
	
	\bibitem{Scolnic:2017caz}
	D.M. Scolnic, et~al., Astrophys. J. \textbf{859}(2), 101 (2018)
	
	\bibitem{Yu:2017iju}
	H.~Yu, B.~Ratra, F.Y. Wang, Astrophys. J. \textbf{856}(1), 3 (2018)
	
	\bibitem{Amirhashchi:2018nxl}
	H.~Amirhashchi, S.~Amirhashchi, Phys. Dark Univ. \textbf{29}, 100557 (2020)
	
	\bibitem{Chen:20db18v}
	L.~Chen, Q.G. Huang, K.~Wang, JCAP \textbf{02}, 028 (2019)
	
	\bibitem{Blake:2011en}
	C.~Blake, et~al., Mon. Not. Roy. Astron. Soc. \textbf{418}, 1707 (2011)
	
	\bibitem{foreman2013emcee}
	D.~Foreman-Mackey, D.W. Hogg, D.~Lang, J.~Goodman, Publ. Astron. Soc. Pac
	\textbf{125}(925), 306 (2013)
	
	\bibitem{newville2021lmfit}
	M.~Newville, R.~Otten, A.~Nelson, A.~Ingargiola, T.~Stensitzki, D.~Allan,
	A.~Fox, F.~Carter, P.D. Micha{\l}, et~al.
	\newblock Lmfit/lmfit-py (2018)
	
	\bibitem{sadri2019observational}
	E.~Sadri, Eur. Phys. J. C \textbf{79}(9), 1 (2019)
	
	\bibitem{Bocquet2016}
	S.~Bocquet, F.W. Carter, Int. J. Open Source Softw. Process. \textbf{1}(6)
	(2016)
	
	\bibitem{Yin:2019rgm}
	Z.Y. Yin, H.~Wei, Eur. Phys. J. C \textbf{79}(8), 698 (2019)
	
	\bibitem{blake2011wigglez}
	C.~Blake, E.A. Kazin, F.~Beutler, T.M. Davis, D.~Parkinson, S.~Brough,
	M.~Colless, C.~Contreras, W.~Couch, S.~Croom, et~al., Mon. Not. Roy. Astron.
	Soc. \textbf{418}(3), 1707 (2011)
	
	\bibitem{pavon2009chatelier}
	D.~Pav{\'o}n, B.~Wang, Gen. Relativ. Gravit. \textbf{41}(1), 1 (2009)
	
	\bibitem{PhysRevD.102.123537}
	C.P. Singh, A.~Kumar, Phys. Rev. D \textbf{102}, 123537 (2020)
	
	\bibitem{Hinshaw2008FiveYearWM}
	G.~Hinshaw, J.~Weiland, R.~Hill, N.~Odegard, D.~Larson, C.~Bennett, J.~Dunkley,
	B.~Gold, M.~Greason, N.~Jarosik, E.~Komatsu, M.~Nolta, L.~Page, D.~Spergel,
	E.J. Wollack, M.~Halpern, A.~Kogut, M.~Limon, S.~Meyer, G.~Tucker, E.~Wright,
	Astrophys. J., Suppl. Ser. \textbf{180}, 225 (2008)
	
	\bibitem{Planck:2018vyg}
	N.~Aghanim, et~al., Astron. Astrophys. \textbf{641}, A6 (2020)
	
	\bibitem{alam2004case}
	U.~Alam, V.~Sahni, A.A. Starobinsky, JCAP \textbf{2004}(06), 008 (2004)
	
	\bibitem{AlMamon:2018uby}
	A.~Al~Mamon, K.~Bamba, Eur. Phys. J. C \textbf{78}(10), 862 (2018)
	
	\bibitem{Mamon:2016wow}
	A.A. Mamon, K.~Bamba, S.~Das, Eur. Phys. J. C \textbf{77}(1), 29 (2017)
	
	\bibitem{Mamon:2016dlv}
	A.A. Mamon, S.~Das, Eur. Phys. J. C \textbf{77}(7), 495 (2017)
	
	\bibitem{Farooq:2016zwm}
	O.~Farooq, F.R. Madiyar, S.~Crandall, B.~Ratra, Astrophys. J. \textbf{835}(1),
	26 (2017)
	
	\bibitem{2003}
	D.N. Spergel, L.~Verde, H.V. Peiris, E.~Komatsu, M.R. Nolta, C.L. Bennett,
	M.~Halpern, G.~Hinshaw, N.~Jarosik, A.~Kogut, M.~Limon, S.S. Meyer, L.~Page,
	G.S. Tucker, J.L. Weiland, E.~Wollack, E.L. Wright, Astrophys. J., Suppl.
	Ser. \textbf{148}(1), 175 (2003)
	
	\bibitem{Valcin:2021jcg}
	D.~Valcin, R.~Jimenez, L.~Verde, J.L. Bernal, B.D. Wandelt, JCAP \textbf{08},
	017 (2021)
	
	\bibitem{Sahni:2002fz}
	V.~Sahni, T.D. Saini, A.A. Starobinsky, U.~Alam, JETP Lett. \textbf{77}, 201
	(2003)
	
	\bibitem{Alam:2003sc}
	U.~Alam, V.~Sahni, T.D. Saini, A.A. Starobinsky, Mon. Not. Roy. Astron. Soc.
	\textbf{344}, 1057 (2003)
	
	\bibitem{gregory_2006}
	R.D. Gregory, \emph{{Classical Mechanics}} (Cambridge University Press, 2006)
	
	\bibitem{Pavon:2012qn}
	D.~Pavon, N.~Radicella, Gen. Rel. Grav. \textbf{45}, 63 (2013)
	
	\bibitem{Crawford:2010rg}
	S.M. Crawford, A.L. Ratsimbazafy, C.M. Cress, E.A. Olivier, S.L. Blyth, K.J.
	van~der Heyden, Mon. Not. Roy. Astron. Soc. \textbf{406}, 2569 (2010)
	
	\bibitem{Carvalho:2011qw}
	J.C. Carvalho, J.S. Alcaniz, Mon. Not. Roy. Astron. Soc. \textbf{418}, 1873
	(2011)
	
	\bibitem{Simon:2004tf}
	J.~Simon, L.~Verde, R.~Jimenez, Phys. Rev. D \textbf{71}, 123001 (2005)
	
	\bibitem{Stern:2009ep}
	D.~Stern, R.~Jimenez, L.~Verde, M.~Kamionkowski, S.A. Stanford, JCAP
	\textbf{02}, 008 (2010)
	
	\bibitem{PhysRevD.74.083520}
	B.~Wang, Y.~Gong, E.~Abdalla, Phys. Rev. D \textbf{74}, 083520 (2006)
	
	\bibitem{egan2010larger}
	C.A. Egan, C.H. Lineweaver, Astrophys. J. \textbf{710}(2), 1825 (2010)
	
	\bibitem{B:2020wuq}
	P.B. Krishna, T.K. Mathew, arXiv:2002.02121
	
	\bibitem{PhysRevD.15.2738}
	G.W. Gibbons, S.W. Hawking, Phys. Rev. D \textbf{15}, 2738 (1977)
	
	\bibitem{2021EPJC...81..644S}
	E.N. {Saridakis}, S.~{Basilakos}, Eur. Phys. J. C \textbf{81}(7), 644 (2021)
	
	\bibitem{2013EPJC...73.2619M}
	T.K. {Mathew}, R.~{Aiswarya}, V.K. {Soman}, Eur. Phys. J. C \textbf{73}, 2619
	(2013)
	
\end{thebibliography}
\end{document}